\documentclass[fleqn,usenatbib]{mnras}

\usepackage{txfonts}

\usepackage[T1]{fontenc}
\usepackage{ae,aecompl}

\usepackage[normalem]{ulem}

\usepackage{graphicx}
\usepackage{tabularx}
\usepackage{subfig}
\usepackage{comment}

\usepackage{amssymb}	
\usepackage{epsfig}
\usepackage[utf8]{inputenc}
\usepackage{dcolumn}
\usepackage{natbib}
\usepackage{enumerate}
\usepackage{longtable,lscape}
\usepackage[usenames,dvipsnames]{xcolor}
\usepackage{grffile}
\usepackage{array,float}
\usepackage{multirow}
\usepackage{lscape}
\usepackage{pdflscape}
\usepackage{afterpage}
\usepackage{capt-of}
\usepackage{xargs}

\usepackage{float}

\usepackage[dvipsnames]{xcolor}
\usepackage{adjustbox}
\usepackage{booktabs}

\def\HII{H{\sc ii}}

\newcolumntype{d}[1]{D{.}{\cdot}{#1}}

\newcolumntype{.}{D{.}{.}{-1}}

\newcommand{\lsun}{L$_\odot$}
\newcommand{\msun}{M$_\odot$}

\newcommand{\mum}{$\mu$m}

\newcommand{\hii}{H{\sc ii}}

\title[Star Formation Efficiencies]{ATLASGAL - Star forming efficiencies and the Galactic star formation rate}

\author[M.\,R.\,A.\,Wells et al.]{M.\,R.\,A.\,Wells$^{1,2}$\thanks{E-mail: {\color{black} wells@mpia.de}},
J.\,S.\,Urquhart$^{1}$\thanks{E-mail: j.s.urquhart@kent.ac.uk}, T.\,J.\,T.\,Moore$^{3}$, K.\,E.\,Browning$^{1}$, S.\,E.\,Ragan$^4$,\newauthor A.\,J.\,Rigby$^4$, D.\,J.\,Eden$^{5}$, M.\,A.\,Thompson$^{6}$\\
$^{1}$ Centre for Astrophysics and Planetary Science, University of Kent, Canterbury, CT2\,7NH, UK \\
$^{2}$ Max Planck Institute for Astronomy, Konigstuhl 17, 69117 Heidelberg, Germany \\
$^{3}$ Astrophysics Research Institute, Liverpool John Moores University, Liverpool Science Park, 146 Brownlow Hill, Liverpool, L3\,5RF, UK\\
$^{4}$ School of Physics and Astronomy, Cardiff University, Queen’s Buildings, The Parade, Cardiff, CF24 3AA, UK\\
$^{5}$ Armagh Observatory and Planetarium, College Hill, Armagh, BT61 9DB, UK\\
$^{6}$ School of Physics and Astronomy, University of Leeds, Leeds LS2 9JT, UK\\
}

\date{Accepted XXX. Received YYY; in original form ZZZ}

\pubyear{2021}

\begin{document}
\label{firstpage}
\pagerange{\pageref{firstpage}--\pageref{lastpage}}
\maketitle

\begin{abstract}
 
The ATLASGAL survey has characterised the properties of approximately 1000 embedded \hii\ regions and found an empirical relationship between the clump mass and bolometric luminosity that covers 3--4 orders of magnitude. Comparing this relation with simulated clusters drawn from an initial mass function and using different star formation efficiencies we find that a single value is unable to fit the observed luminosity to mass ($L/M$) relation. We have used a  Monte Carlo simulation to generate 200,000 clusters using the $L/M$-ratio as a constraint to investigate how the star formation efficiency changes as a function of clump mass. This has revealed that the star formation efficiency decreases with increasing clump mass with a value of 0.2 for clumps with masses of a few hundred solar masses and dropping to 0.08 for clumps with masses of a few thousand solar masses. We find good agreement between our results and star formation efficiencies determined from counts of embedded objects in nearby molecular clouds. Using the star formation efficiency relationship and the infrared excess time for  embedded star formation of $2\pm1$\,Myr we estimate the Galactic star formation rate to be approximately $0.9\pm0.45$\,\msun\,yr$^{-1}$, which is in good agreement with previously reported values. This model has the advantage of providing a direct means of determining the star formation rate and avoids the difficulties encountered in converting infrared luminosities to stellar mass that affect previous galactic and extragalactic studies.

\end{abstract}




\section{Introduction}
\label{sect:intro}

High-mass stars ($> 8$\,\msun) drive many of the fundamental processes in astrophysics.
They are responsible for the chemical enrichment of the interstellar medium (ISM), and for injecting the radiative and mechanical energy \citep{2007ARA&A..45..481Z}. On local scales they regulate star formation and determine the initial conditions for the formation of planetary systems \citep{2018sf2a.conf..311L}, while on the larger scale, they drive the evolution of the galaxies themselves (\citealt{kennicutt2012}).

Most star formation takes place in giant molecular clouds (GMCs), these have sizes of several tens of parsecs and contain millions of Solar masses of interstellar gas and dust (\citealt{heyer2015}). GMCs have a hierarchical structure consisting of diffuse material and dense clumps within which denser cores can be found. The star formation process starts when the clumps become unstable to gravity and begin to collapse, fragmenting into cores that subsequently collapse into an individual protostar or small group of protostellar objects. The protostars continue to evolve acquiring more mass via disc accretion (\citealt{Motte_2018}). 

Unlike their lower-mass counterparts, high-mass stars reach the main sequence while still deeply embedded in their natal clump and so all of the earliest stages occur behind many hundreds of magnitudes of visual extinction, rendering these stages invisible even at mid-infrared wavelengths. To complicate matters further, high-mass stars are rare and evolve quickly, resulting in very few being available to study during their formation, especially since high-mass star forming regions are much more distant than low-mass counterparts. These difficulties have traditionally hindered our understanding of high-mass star formation; however, this has improved dramatically over the past 20 years with the commissioning of new facilities such as ALMA, APEX, the \textit{Spitzer} and \textit{Herschel} space telescopes and upgrades to existing facilities such as the VLA and NOEMA. These facilities have driven a slew of surveys of the Galactic plane covering the near-infrared to radio wavelengths. These surveys have provided the high resolution and large volumes needed to obtain large and statistically representative samples of high-mass star forming clumps. Studying the evolution of star formation taking place within a large sample of clumps will provide insight into how efficiently molecular gas can be converted into stellar clusters and what impact environment plays in this process.  

The discovery of a strong correlation between dense gas and the star formation rate (SFR; mass of star produced per unit time) extends from extragalactic scales (e.g. \citealt{kennicutt1998, bigiel2008}) to clump scales (e.g. \citealt{wu2005,lada2010,heiderman2010,wu2010}). This means that understanding the star formation in the Galaxy will provide a framework for interpreting star formation in nearby and in high-$z$ galaxies (\citealt{heiderman2010}). However, comparisons of star formation rates of Galactic clumps using YSO source counts and stellar mass derived from infrared luminosities has revealed a significant difference, with the latter being found to underestimate the SFR by up to an order of magnitude (e.g. \citealt{heiderman2010, gutermuth2011}). The largest differences are mostly due to the CO being used to determine masses of galaxies, which significantly overestimates the mass involved in star formation (e.g. \citealt{gao2004a, gao2004b}), however, even when only the dense gas is considered, the infrared luminosities derived SFR is a factor of $\sim$2 lower than derived from YSO counts (\citealt[][]{2011Chomiuk, Faimali2012}).

In this paper we develop a model that avoids these particular problems. The model creates synthetic clusters by drawing stars from an initial mass function and deriving the luminosity from the stellar mass. We use the results from the APEX Telescope Large Area Survey of the Galaxy (ATLASGAL; \citealt{schuller2009}) survey to investigate the star formation efficiency (SFE; ratio of the amount of dense material converted into stars) using an empirical relation derived from the ATLASGAL survey and a simple model to produce thousands of synthetic clusters. We will use the results of this analysis to estimate the Galactic star formation rate and compare this to previously reported values. 

The structure of the paper is as follows: an overview of the sample being used and details on how the cut-off point for the simulation is calculated is given in Sect.\,\ref{sect:atlasgal}. In Sect.\,\ref{sect:sfe}, we describe how the simulation works and the assumptions we have used and present our theoretical star formation efficiency relationship. In Sect.\,\ref{sect:SFR} we use the star formation efficiencies to obtain estimates for the Galactic star formation rate. We discuss the assumptions used in Sect.\,\ref{sect:discussion} to evaluate our results and outline our main conclusions in  Sect.\,\ref{sect:conclusions}.

\section{ATLASGAL}
\label{sect:atlasgal}

\begin{figure}
    \centering
    \includegraphics[width = 0.49\textwidth, trim=20 0 20 0]{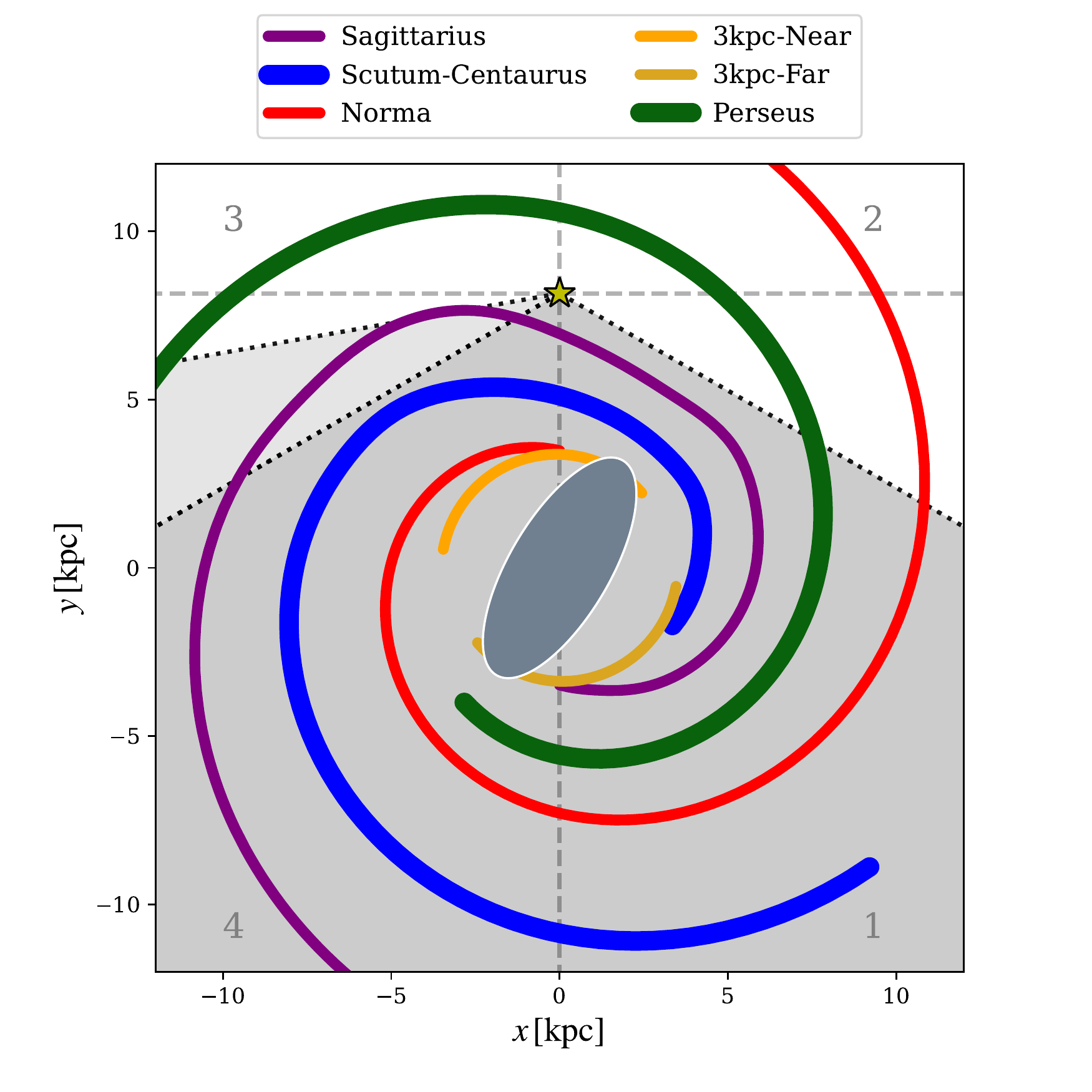}
    \caption{Schematic showing the loci of the spiral arms according to the model by \citet{taylor1993} and updated by \citet{cordes2004}, with an additional bisymmetric pair of arm segments added to represent the 3\,kpc arms. The dark grey shaded region indicates the coverage of the main ATLASGAL survey while the light grey shaded region shows the longitude coverage of the ATLASGAL extension (see text for details). The star shows the position of the Sun and the numbers identify the Galactic quadrants. The bar feature is merely illustrative and does not play a role in our analysis.}
    \label{fig:atlasgal}
\end{figure}

The sample of sources used in this project is drawn from the ATLASGAL \citep{schuller2009,beuther2012,csengeri2014} survey. The survey was conducted using the APEX telescope and is an unbiased 870-$\mu$m continuum survey with an angular resolution of 19\,arcseconds. ATLASGAL initially concentrated on the inner part of the Galactic plane ($300^\circ < \ell < 60\degr$ and $|\,b\,| < 1.5\degr$) but was later extended to include the Sagittarius tangent in the fourth quadrant ($280^\circ < \ell < 300\degr$ and $-2\degr < |\,b\,| < 1\degr$). ATLASGAL covers approximately two thirds of the Galactic molecular disc including all of the molecular gas within the Solar circle ( $R_{\rm gc} < 8.35$\,kpc; \citealt{urquhart2014_atlas}) and significant parts of all of the spiral arms (as shown in Fig.\,\ref{fig:atlasgal}).

The primary goal of ATLASGAL is to provide a large sample of massive dense clumps in the Galaxy that is representative of the early evolutionary stages of high mass star formation \citep{schuller2009}. A catalogue of approximately 10,000 dense clumps has been constructed from the survey data (\citealt{contreras2013,urquhart2014_atlas}) with diameters of $\sim$1\,pc and masses of several hundred \(\textup{M}_\odot\). All of these clumps fulfil the column density threshold for efficient star formation derived by \citet{lada2010} and \citet{heiderman2010} (i.e. 116-129\,\msun\,pc$^{-2}$) and the majority satisfy the \citet{kauffmann2010a} size-mass criterion for high-mass star formation (see \citealt{urquhart2018} for details). This sample of clumps consists of sites of current and future star formation within the inner Galaxy. Given their sizes and masses, these clumps are likely to form stellar clusters.

We have used results from radio and mid- and far-infrared continuum studies (e.g., CORNISH: \citealt{purcell2013,hoare2012}; the RMS survey: \citealt{lumsden2013}; GLIMPSE: \citealt{2003PASP..115..953B}; Hi-GAL: \citealt{Molinari2010}) to produce a well-characterised sample of high-mass star-forming clumps (\citealt{urquhart2018}). We have separated this sample into four distinct  stages (quiescent, protostellar, young stellar object (YSO) and ultra-compact (UC) \hii\ regions) and have shown that these stages are consistent with an evolutionary sequence with increasing temperatures, luminosities and luminosity-to-mass ratios (\citealt{urquhart2022}). This is the largest and most well-characterised sample of high-mass star forming clumps constructed to date.


\begin{figure*}
    \centering
     \includegraphics[width = 0.9\textwidth]{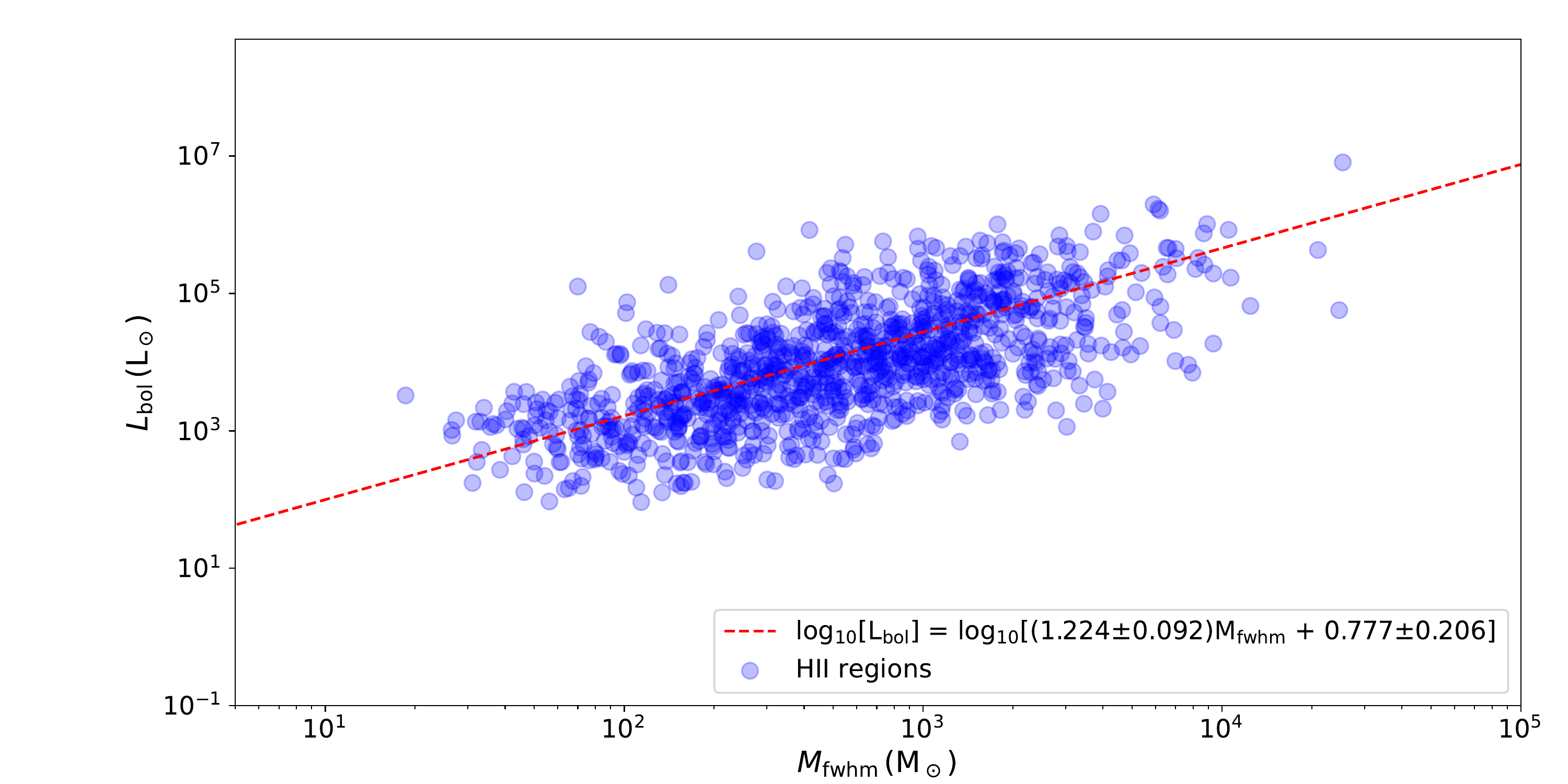}

    \caption{Luminosity-mass plot for all \HII\ regions associated with ATLASGAL clumps. The dashed red line shows the linear least-squares fit to the distribution of \hii\ regions (\citealt{urquhart2022}).}

    \label{fig:lm_diagram_hii}
\end{figure*} 

\section{Star Formation Efficiency}
\label{sect:sfe}

The UC \hii\ region stage represents an interesting phase in the evolution of high-mass star formation as this is when accretion is halted and the final mass of the star is set, but before the natal material is dispersed and the stars become observable at optical wavelengths. Given that the vast majority of stars form in clusters (\citealt{carpenter2000, lada2003}), including nearly all high-mass stars, it is safe to assume that the \hii\ regions that we have identified are powered by the most massive star in a young protocluster and that the total luminosity is that of all of the stars in the cluster (\citealt{wood1989b,walsh2001}). These \hii\ regions, therefore, represent a stage in which the luminosity is at a maximum and they are still embedded in their natal clump.

For the ATLASGAL clumps, masses have been determined from the integrated 870-\mum\ flux, the dust temperature (\citealt{konig2017}) and the distance, 
available from the ATLASGAL catalogue (\citealt{urquhart2022}). The total luminosity has been determined from greybody fits to the spectral energy distribution constructed from mid-infrared and submillimetre photometry (\citealt{konig2017,urquhart2018}).

In Figure\,\ref{fig:lm_diagram_hii} we show the luminosity-mass relation of the ATLASGAL clumps associated with UC \hii\ regions. This plot reveals a strong correlation between the clump mass and the bolometric luminosity (Spearman correlation coefficient $r_s = 0.66$ with a $p$-value $\ll 0.0013$), which extends over four orders of magnitude in mass and five orders of magnitude in luminosity. This has been noted in previous papers in the series (e.g., \citealt{urquhart2014_csc, urquhart2018, urquhart2022}) and also in previous studies that have included a significant number of embedded \hii\ regions (e.g., \citealt{mueller2002} and \citealt{sridharan2002}). The dashed red line shows the fit to the distribution and marks the transition between the \emph{main accretion} and \emph{envelope clean-up} phase (\citealt{molinari2008}). Given that the \hii\ regions identified by ATLASGAL represent the end of star formation and the beginning of the dispersion of any remaining molecular gas, this empirical relation provides a measure of the maximum luminosity an embedded proto-cluster can attain as a function of the ATLASGAL clump mass and has the form: log($L_{\rm cluster}$) = 1.22 $\pm 0.09$\,log($M_{\rm fwhm}$), where $M_{\rm fwhm}$ is the mass within the 50\,per\,cent 870\,\mum\ flux contour (see \citealt{urquhart2022} for details). 

By linking the cluster luminosity to the cluster mass through an initial mass function (IMF; e.g. \citealt{kroupa2001}) we can calculate the star formation efficiency of clumps ($\varepsilon$) using (\citealt{lada2003}):

\begin{equation}
\varepsilon = \frac{M_\mathrm{stars}}{M_\mathrm{gas} + M_\mathrm{stars}},
   \label{eq:sfe1}
\end{equation}

\noindent where \emph{$M_\mathrm{stars}$} is the total mass of stars in the cluster and \emph{$M_\mathrm{gas}$} is the gas mass remaining after the star formation has ceased.

To calculate the star formation efficiency we need to be able to convert the cluster luminosity into a cluster mass. To do this we make the following three assumptions that allow us to link the total luminosity measured to a range of cluster masses (since there are a number of ways you can construct a cluster that is consistent with the IMF and total luminosity):

\begin{itemize}
    \item The distribution of stellar masses within each cluster is dictated by the initial mass function (IMF; \citealt{Salpeter}), which describes the relative number of stars as a function of their initial mass.\\
    
    \item The range of stellar masses goes from 0.1 to 120 \(\textup{M}_\odot\) (\citealt{kroupa2001}); less than 0.1 \(\textup{M}_\odot\) the core becomes degenerate before the temperature rises enough for fusion to begin, and larger than 120 \(\textup{M}_\odot\), the stability and equilibrium of the star will start to be compromised.\\
    
    \item The mass of the clumps is relatively constant during the star formation process with mass loss due to outflows being replaced by infalling material (see Sect.\,\ref{sect:discussion} for more on this point) and/or the SFE is low.
    
\end{itemize}

To determine the range of possible cluster masses we have followed the procedure described by \citet{walsh2001} using a Monte Carlo method to randomly sample from an IMF (see also \citealt{sridharan2002}). This function is most commonly expressed as a power law and defined over a large range of stellar masses (\citealt{2007prpl.conf..149B}). The two most commonly used IMFs are those of \citet{Chabrier2003} and \citet{kroupa2001} and both have similar distributions for stars with masses above 0.1\,\msun\ and so the choice is not critical, however, the latter is simpler to add into our model and is therefore used in this work (i.e. $N \propto M^\alpha$, where $\alpha = 1.3$ for 0.08\,\msun\, $\le M \le $ 0.5\,\msun\ and $\alpha = 2.3$ for $M > 0.5$\,\msun). 

\begin{figure}
 
  \includegraphics[width=\linewidth]{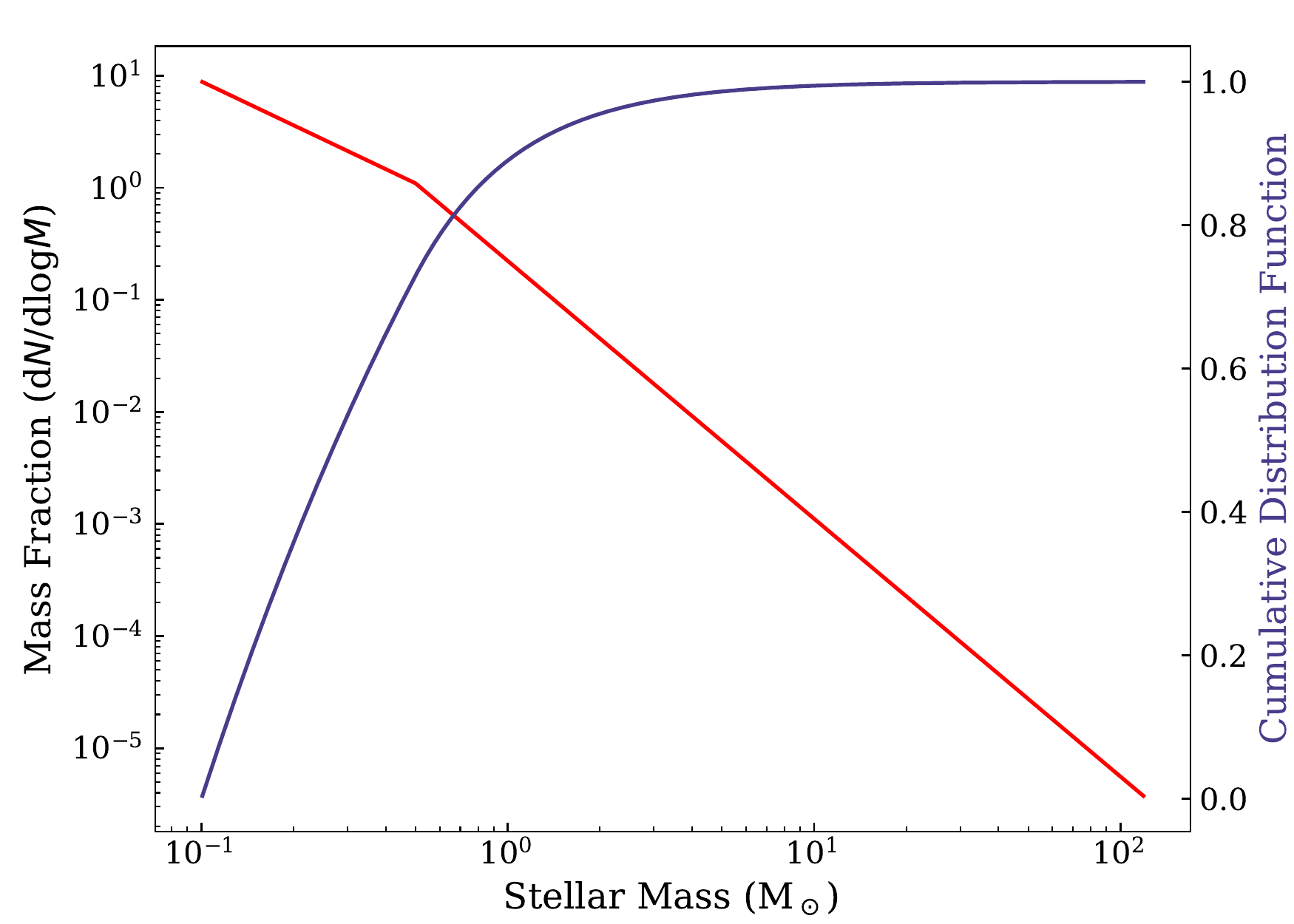}
  \includegraphics[width=\linewidth]{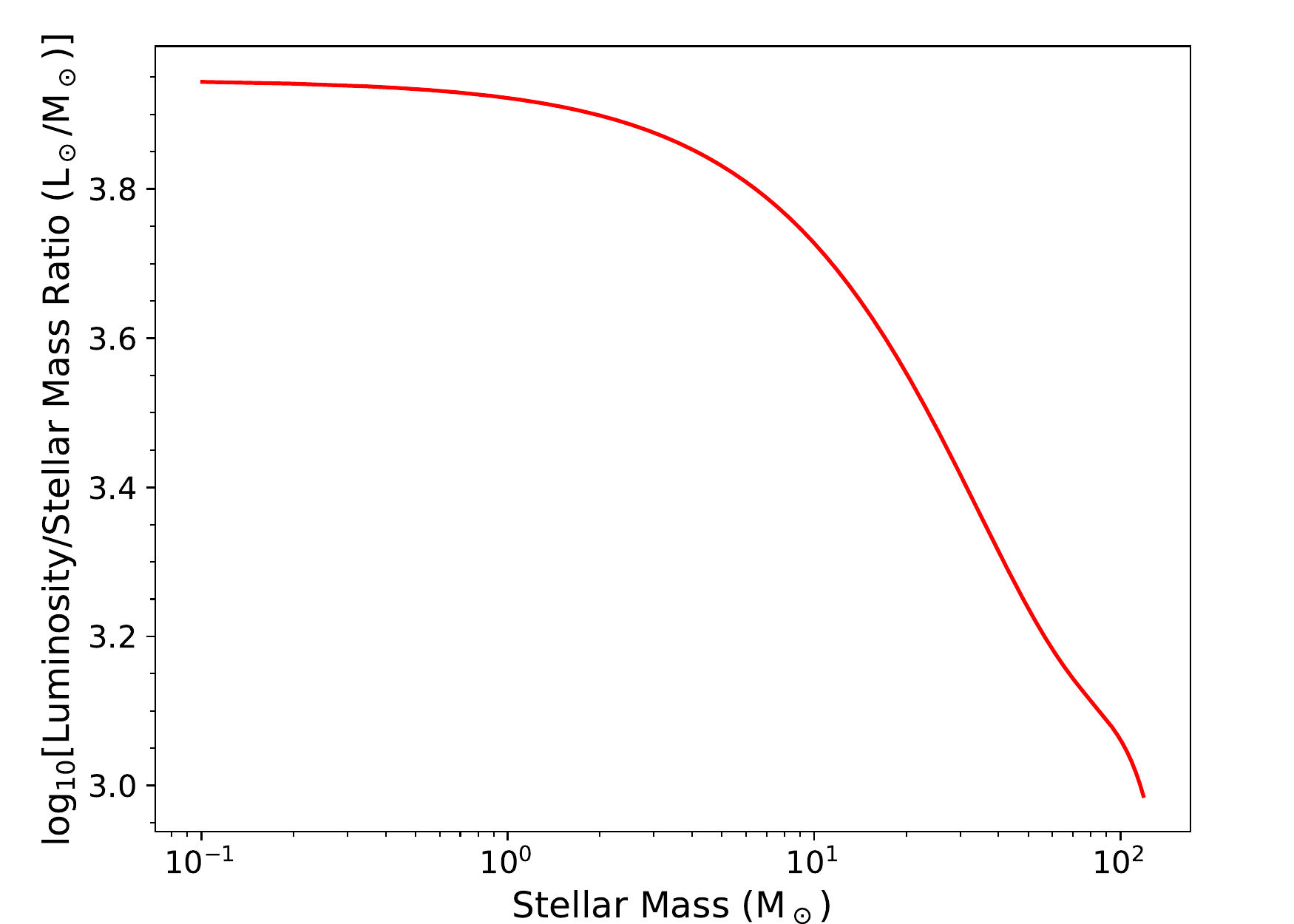}
  \caption{The top panel shows the Kroupa IMF \citep{kroupa2001} (red line) and the corresponding cumulative distribution function showing the probability of sampling a star within a particular mass interval (blue line). The lower panel shows the luminosity to mass ratio for main-sequence stars.
}  
  \label{fig:kroupa}

\end{figure}

Starting with a mass of 0.1\,\msun\ and increasing in steps of 10$^{0.003}$ results in 1026 bins between the range of 0.1 to 120\,\msun. These mass intervals are multiplied by the fraction of the population dictated by the IMF before being summed up and normalised to produce a cumulative distribution function of stellar masses. The \citet{kroupa2001} IMF and the corresponding cumulative distribution produced from the two steps described above are shown in the upper panel of Fig.\,\ref{fig:kroupa}. The cumulative distribution function shows the random probability of selecting a star of a particular mass from the given IMF.  

 We start the process of producing a synthetic cluster by generating a random number between 0 and 1 to select a position on the y-axis of the cumulative distribution shown in the upper panel of Figure\,\ref{fig:kroupa} to determine the mass of a star. We use the luminosity-to-mass relation for main sequence stars (i.e., \citealt{davies2011}) to determine the star's luminosity (shown in the lower panel of Fig.\,\ref{fig:kroupa}). The star's mass and luminosity are added to the cluster properties, then a second random number is generated and used to determine the mass and luminosity of the next cluster member whose properties are added to those of the cluster, and so on and so forth. This process continues until a termination criterion is reached (this can be a given cluster mass, luminosity or star formation efficiency).

 \begin{figure}
    \centering
    \includegraphics[width=.4\textwidth]{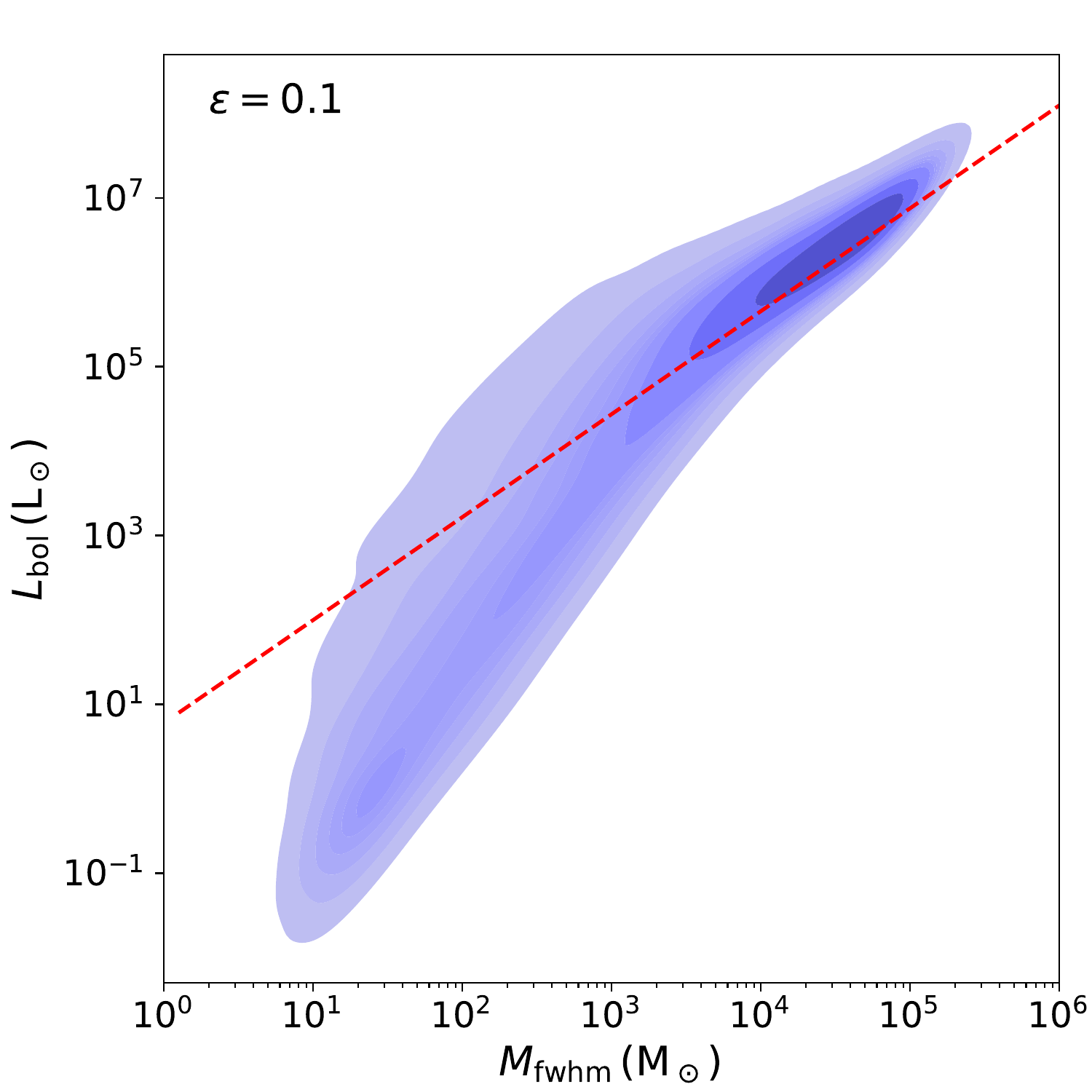}\\
    \includegraphics[width=.4\textwidth]{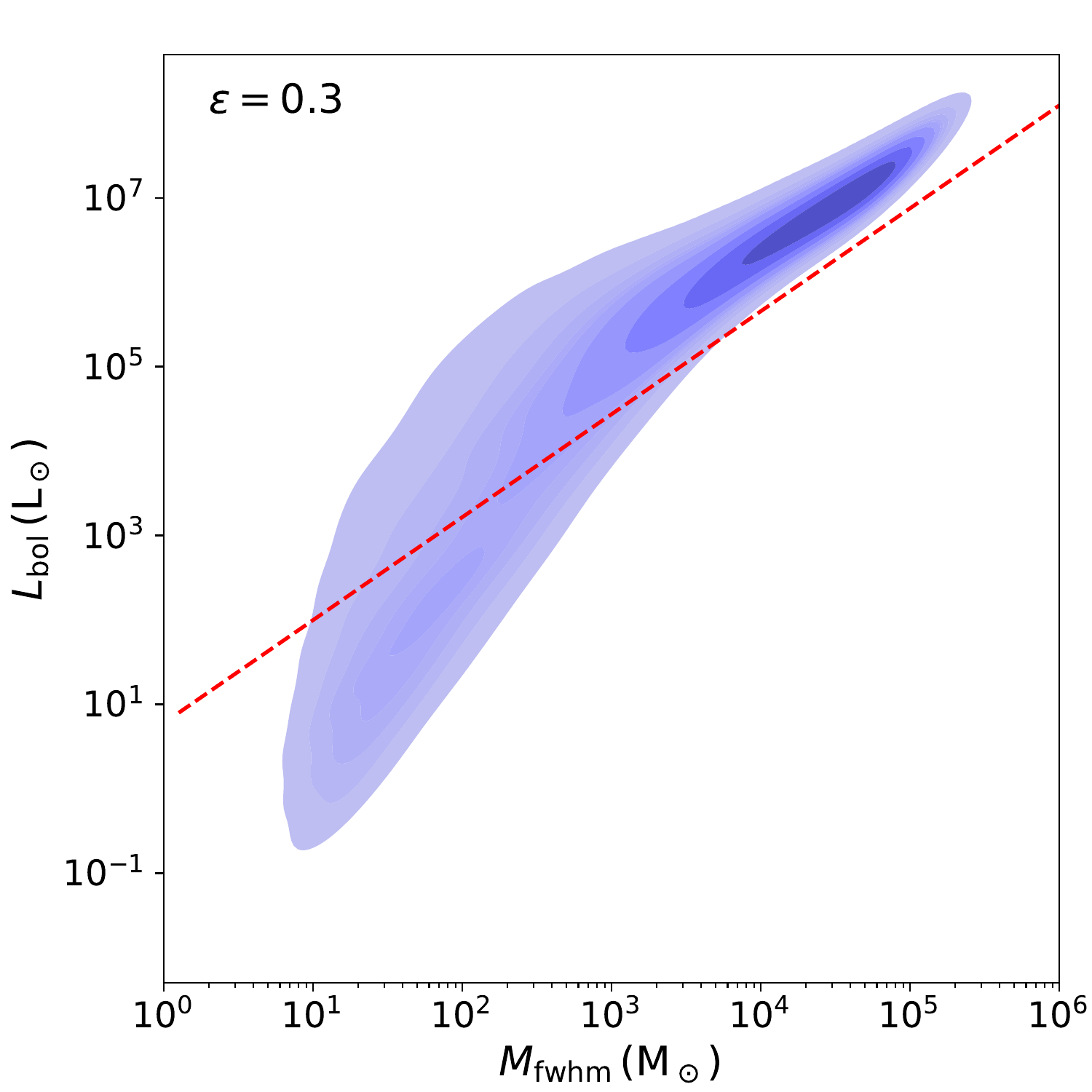}
    \includegraphics[width=.4\textwidth]{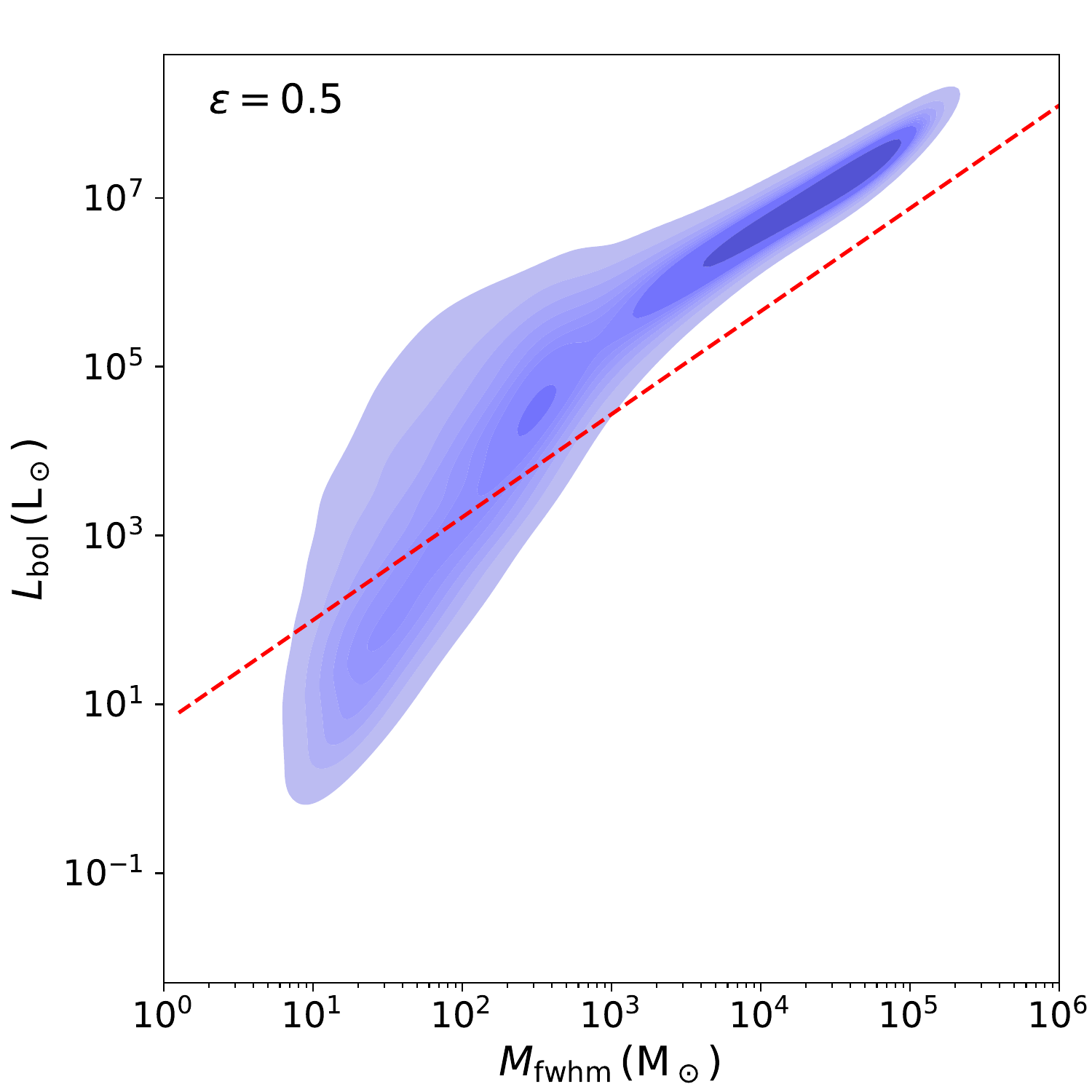}
    
    \caption{Luminosity-mass plots showing the probability
    distribution for the synthetic clusters determined using a standard IMF  \citep{kroupa2001} and assuming star formation efficiencies of 0.1, 0.3 and 0.5 (upper, middle and lower panels, respectively). The dashed red line shows the linear least squares fit to the distribution of \hii\ regions shown in Fig.\,\ref{fig:lm_diagram_hii}.
    }

    \label{fig:model_lm_for_clusters}
\end{figure}

\begin{figure*}
    \centering
    \includegraphics[width = 0.95\textwidth]{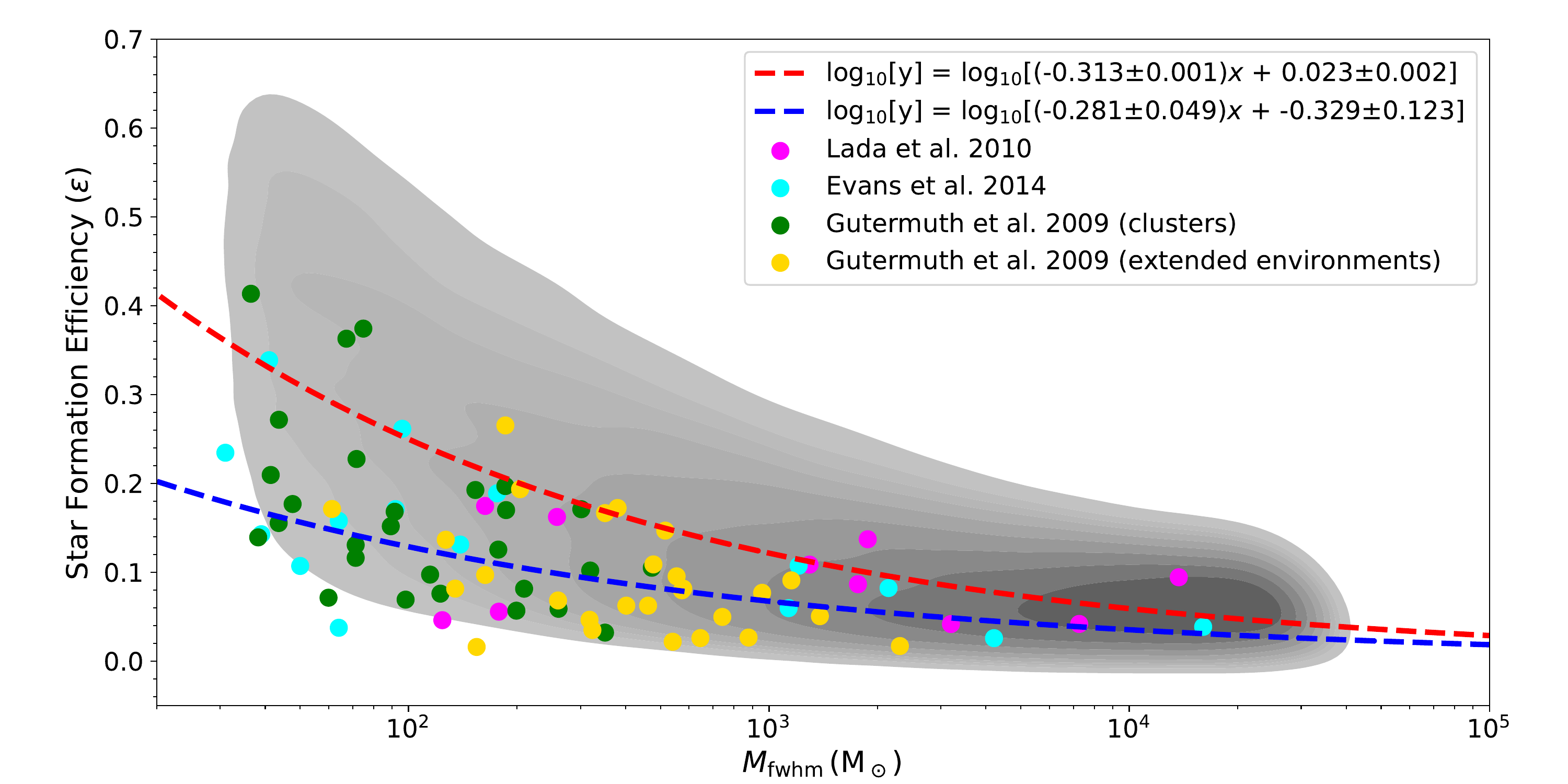}
    \caption{Distribution of star formation efficiencies obtained from the Monte Carlo code as a function of clump mass. The contours show the probability density of simulated clusters across the parameter space. The lowest contour starts at 10\,per\,cent and the highest at 90\,per\,cent. The circles indicate star formation efficiencies available in the literature for nearby molecular clouds ($ < 1.7$\,kpc) that have been determined by counting the number of embedded YSOs. The red and blue dashed lines are power-law fits to the model and observational data respectively.} 
    \label{fig:sfe_dist_map}
    
\end{figure*}

 This method can be used to create synthetic clusters of arbitrary size that are consistent with the IMF, however, we need to constrain the total mass. The simplest way to do this is to use the ATLASGAL clump mass as a reservoir and fix the star formation efficiency ($\varepsilon$).  We build these clusters by successively adding stars until the total mass exceeds the required star formation efficiency, at which point the cluster is considered complete and the cluster luminosity is recorded. This process was repeated 200\,000 times for clump masses between 30 to $5\times 10^4$\,\msun. 
 
 In Figure\,\ref{fig:model_lm_for_clusters} we show the range of bolometric luminosities for a given clump mass for synthetic clusters produced using the three different values for the star formation efficiency ($\varepsilon$ values of 0.1, 0.3 and 0.5). These plots also include the empirical relationship between mass and luminosity obtained from the UC \hii\ regions from Fig.\,\ref{fig:lm_diagram_hii}. Comparing the fit from the data with the distributions of the synthetic clusters (blue contoured regions) reveals that the lower value of $\varepsilon = 0.1$ provides a good fit to the data for the higher-mass clumps ($\gtrapprox 1000$\,\msun); however, for lower-mass clumps a higher value of  $\varepsilon$ provides a better fit ($\varepsilon = 0.3$ for  100\,\msun\ $< M_{\rm fwhm} <$ 1000\,\msun\ and $\varepsilon > 0.3$ for $M_{\rm fwhm} <$ 100\,\msun). This indicates that there is a relationship between clump mass and star formation efficiency resulting in $L(M) \sim M$ rather than the expected $\sim M^2$ for main-sequence clusters, which suggests that the SFE decreases with increasing clump mass. Such a tendency was first noted in \citet{urquhart2013_cornish}.

We can explore this hypothesis in a little more detail by rerunning the Monte Carlo code, but rather than terminating the cluster-building when a specified star formation efficiency threshold is reached, we use the empirical relationship determined from the \hii\ regions derived earlier in this section (see also Fig.\,\ref{fig:lm_diagram_hii}). We can then use the range of cluster masses to determine how the star formation efficiency changes as a continuous function of mass in a more useful way.

In Figure\,\ref{fig:sfe_dist_map} we show the results of such Monte Carlo modelling and a power-law fit that quantifies the relationship, which is given below in terms of the physical parameters:

\begin{equation}
   {\rm log}_{\rm 10}[\varepsilon]  = {\rm log}_{10}[mx +c].
   \label{eq:sfe}
\end{equation}

\noindent where $x = {\rm log_{10}}[M_\mathrm{fwhm}]$ in solar masses and the values of the coefficients $m$ and $c$ are $-0.313\pm0.001$ and $0.022\pm0.002$ respectively. The star formation efficiency is estimated from the cluster mass just before and after the luminosity threshold is breached.

Although the spread of allowed values is relatively large, this plot clarifies the trend for decreasing star formation efficiency as a function of increasing clump mass hinted at in Fig.\,\ref{fig:model_lm_for_clusters}. The average star formation efficiency is 0.20 for clumps less than $\sim$500\,\msun, dropping to 0.15 for clumps between 500 and a few thousand \msun, before falling to $\sim$0.08 for more massive clumps. Given that star formation efficiencies have been calculated from clumps hosting \hii\ regions, the expansion of which is likely to disperse the natal material, and thus ending star formation within the clump, these can be considered to be upper limits to the possible star formation efficiencies. 

There are a number of measurements of the star formation efficiency available in the literature we can use to compare with the results obtained from our simple model. We have selected three studies of nearby molecular clouds ($<$ 1.7\,kpc; \citealt{lada2010,2014Evans,gutermuth2009}) that have measured the SFE by counting the number of embedded YSOs identified in mid-infrared maps from the {\em Spitzer} surveys and gas masses from near-infrared extinction maps. 
\citet{lada2010} compiled a sample of 11 local molecular clouds from the literature located within 0.5\,kpc. We have derived the SFE for these clouds using Eqn.\,\ref{eq:sfe1} by first estimating the total mass of stars assuming an average stellar mass of 0.5\,\msun\ (e.g., \citealt{muench2007}) and using the given dense-gas mass, which they define as gas with an extinction of $A_K > 0.8\pm0.2$\,mag, which corresponds to a visual extinction ($A_V$) of $7.3\pm1.8$\,mag. 

\citet{2014Evans} provides a table of parameters for 29 molecular clouds compiled from the c2d (\citealt{evans2003,evans2009}) and Gould Belt (\citealt{dunham2013}) {\em Spitzer} legacy programmes (typically  located within a few 100 pc). We estimate the SFE from these results by converting the star formation rate, given in solar masses per million years (\msun\,Myr$^{-1}$), into YSO mass assuming a formation time of 2\,Myr (e.g. \citealt{covey2010}) and dividing this by the cloud's mass of dense gas and YSO stellar mass (defined as the mass above an extinction contour of $A_V = 8$\,mag).
Here, we only consider clouds with a dense mass larger than 30\,\msun\ in order to be consistent with the model; this reduces the sample from 29 to 18 clouds. 

The final sample we look at is a study by \citet{gutermuth2009} that concentrates on the clump properties and those of their embedded cluster population and is considered to provide a good comparison with the ATLASGAL catalogue. They provide stellar surface densities and a measurement of the mean $A_{\rm K}$ extinction for clusters and their more extended environment; these can be used to estimate the SFE. We convert the extinction into a column density using $N(\rm H_2) = 1.111 \times 10^{20} A_{\rm K}$\,cm$^{-2}$
(\citealt{lacy2017}),
and then into a mass surface density using the following relation: (c.f. \citealt{bohlin1978, gutermuth2011, pokhrel2020}):

\begin{equation}
\Sigma_{\rm gas} = \overline{N(H_2)}\,\frac{15}{0.94\times 10^{21}}\,{\rm M_\odot\,pc^{-2}}.
\end{equation}

We estimate the SFE of the clouds and clumps by dividing the surface density of embedded protostars by the mass surface density. To ensure that we are comparing similar density structures we restrict the \citet{gutermuth2009} sample to those with mean $A_K \geq 0.5$, which corresponds to the nominal detection threshold of the ATLASGAL survey ($3.6\times 10^{21}\,{\rm cm^{-2}}$ and $1\times 10^{22}\,{\rm cm^{-2}}$ for dust temperatures of 30\,K and 10\,K respectively; \citealt{schuller2009_full}); 24 extended regions and 38 clusters satisfy this criterion.

The $A_V$ threshold of $\sim$8 used by these studies corresponds to a volume density $n_{H_2} \geq 10^4$\,cm$^{-3}$, which compares well with the volume density determined for ATLASGAL clumps ($n_{H_2} \approx 10^{4.5 \pm0.5}$\,cm$^{-3}$; \citealt{urquhart2022}), therefore making them ideal for comparison. Comparing the distribution of these observational measurements with the star formation efficiency predicted by our simple model reveals them to be in very good agreement. The 4 studies provide a total of 91 SFE measurements, 83 of which fall within the envelope of values predicted by the model, corresponding to $\sim$90\,per\,cent agreement. The clouds that fall outside the predicted range tend to be at the lower-mass end of the sample where the star formation may not be well represented by the chosen IMF or may be examples of clouds in a very early stage of star formation. Indeed, this sample includes the Pipe Nebula and the Lupus 4 molecular clouds, which are associated with the lowest levels of star formation activity, with only 21 and 12 YSOs, respectively \citep{lada2010}. However, all but two of the observational measurements that fall just outside the 10\,per\,cent probability contour are within the 5\,per\,cent contour and therefore within 2$\sigma$ of the mean value.

Turning our attention back to the SFE measurements that do fall into our model probability distribution, we note that they have a similar distribution in parameter space (i.e. large variation in SFE for lower mass clumps and much narrower range for higher mass clumps) but that the majority are located below the fit in Fig.\,\ref{fig:sfe_dist_map}, 
indicating that the measured values are not so consistent with the model predictions. We also perform a power-law fit to the survey data to quantify the difference between the model and the observational data and this is shown as a dashed blue line in Fig.\,\ref{fig:sfe_dist_map}; this reveals a difference of a factor of $\sim$1.75 for clumps mass of $\sim$1000\,\msun, rising to 1.9 for lower mass clumps ($\sim 100$\,\msun) and decreasing to 1.6 for high mass clumps ($\sim 10^4$\,\msun). However, there are a few differences between the model and the data that can resolve this issue. First, the model is based on the $L/M$-ratio for \hii\ regions, where the star formation has either ended or is coming to an end and the SFE is at its maximum value. The SFE measurements are for nearby molecular clouds covering the complete range of time-scales for star formation, from clouds where the star formation is only just beginning, to clouds where it is coming to an end; these are {\em instantaneous} SFE measurements. So, in general, we should expect these to be lower than predicted by our model, which is estimating the final SFE of dense clumps. Second, the SFE measurements are produced by counting the number of mid-infrared point sources seen in {\em Spitzer} images and so are only sensitive to mid-infrared-bright YSOs and are likely to miss some lower-luminosity YSOs and young protostellar objects that only become visible at longer wavelengths. Furthermore, {\em Spitzer} requires IR-excess emission in order to identify cluster members but disc fractions are of order 50\,per\,cent and correcting the source counts for the missing discless sources would be a plausible way to solve the apparent data-model discrepancy. The SFE measurements may be incomplete and, therefore, underestimate the current level of star formation. Another possibility that the model is based on the luminosity-mass relation for \hii\ regions and this may not be appropriate for low-mass star formation regions. 

\begin{figure*}
    \centering
    \includegraphics[width = 0.9\textwidth]{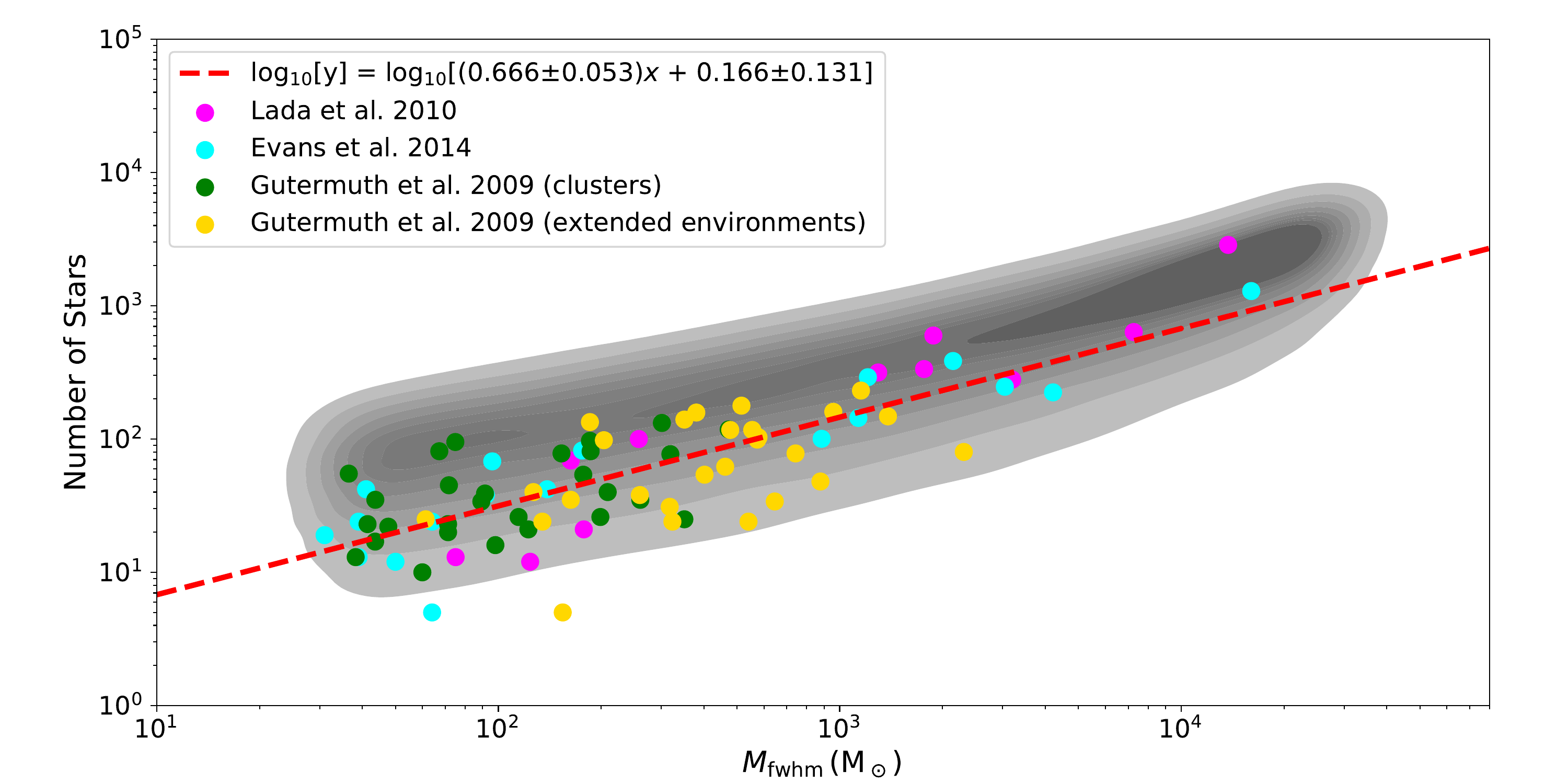}
    \caption{Comparison of YSO number counts from studies of nearby molecular clouds and clumps compared to the number of stars in each of the simulated clusters. The lowest contour starts at 10\,per\,cent and the highest at 90\,per\,cent. 
    The dashed red line is a linear fit to the log values of clump mass and star counts from the combined literature data (see text for discussion).}
    \label{fig:NYSO_histo}
    
\end{figure*}

In Figure\,\ref{fig:NYSO_histo} we compare the number of stars detected in nearby molecular clouds to the stellar membership predicted by our Monte Carlo simulation. On this plot we also show the results of a linear least squares fit to the log values of the ensemble of observational measurements (dashed red line), this has a slope of $0.67\pm0.05$, which is significantly shallower than the slope of $0.96\pm0.13$ obtained by \citet{lada2010}, but in general agreement given the uncertainties associated with both measurements.\footnote{We use a linear fit here to allow comparison with the results of \citet{lada2010}.} This is also broadly consistent with model results. \citet{2014Evans} investigated the star formation in a sample of nearby clumps and massive dense clumps and reported a slope of 0.89 from a fit to the logs of the dense gas mass and SFR (see also \citealt{wu2005,vutisalchavakul2013}). The number of stars forming in the massive dense clumps is determined by integrating the radio continuum emission over the whole of the associated \hii\ region, while the mass is estimated from the HCN (1-0) molecular line transition by \citet{wu2010}. However, as \citet{2014Evans} points out, the masses estimated for the massive dense clumps are likely to be much lower than initial values, due to dispersion of material by the \hii\ region; if this was taken into account the slope would be significantly shallower, which would bring it more in line with the slope determined by our model.

Figures\,\ref{fig:sfe_dist_map} and \ref{fig:NYSO_histo} demonstrate the good agreement between the model and observational measurement. The good correlation between the model and the observations is also notable as it demonstrates that the model is able to determine reliable SFRs for more distant marginally resolved clumps where counting YSOs is not possible. It also avoids the issue that has plagued previous attempts to estimate the SFRs from infrared luminosities that underestimate the SFR by a factor of 2-3 (as discussed in Sect.\,\ref{sect:intro}). The consistency between our model's predictions and the values available in the literature give us confidence that our model is reliable and can be applied to the ATLASGAL clumps to predict their future star formation efficiency and contribution to the Galactic star formation rate. We will focus on this in the next section.

\section{Galactic Star formation rate}
\label{sect:SFR}

In the previous section we have obtained  a theoretical relation between the clump mass and star formation efficiency. This allows us to predict how much of the dense gas is likely to be converted into stars over the star formation time-scale. In this section we will use this relation and a few additional assumptions to estimate the star formation rate for each clump and, by integrating these, obtain an estimate for the Galactic star formation rate. 

The star formation rate describes how quickly material is converted into stars and is given by:

\begin{equation}
    {\rm SFR} =\frac{\varepsilon \times M_\mathrm{fwhm}}{\tau_{\rm sf}},
\end{equation}

\noindent where $\tau_{\rm sf}$ is the star formation time-scale. We have already determined the clump masses and a relationship for the star formation efficiency and we just need to select a suitable value for the star formation time scale.

A realistic choice for the star formation time-scale comes from work by \citet{covey2010} who estimate the duration of infrared excess to be $2\pm1$\,Myr
for embedded YSOs ($\tau_{\rm YSO}$). This is the value used by the three studies discussed in the previous section to estimate the star formation rates. This is reasonable, given that these clouds are predominately forming low-mass stars where the radiative and mechanical feedback is unlikely to have a significant impact on the structure of the cloud itself. 

\begin{figure}
  \includegraphics[width=\linewidth]{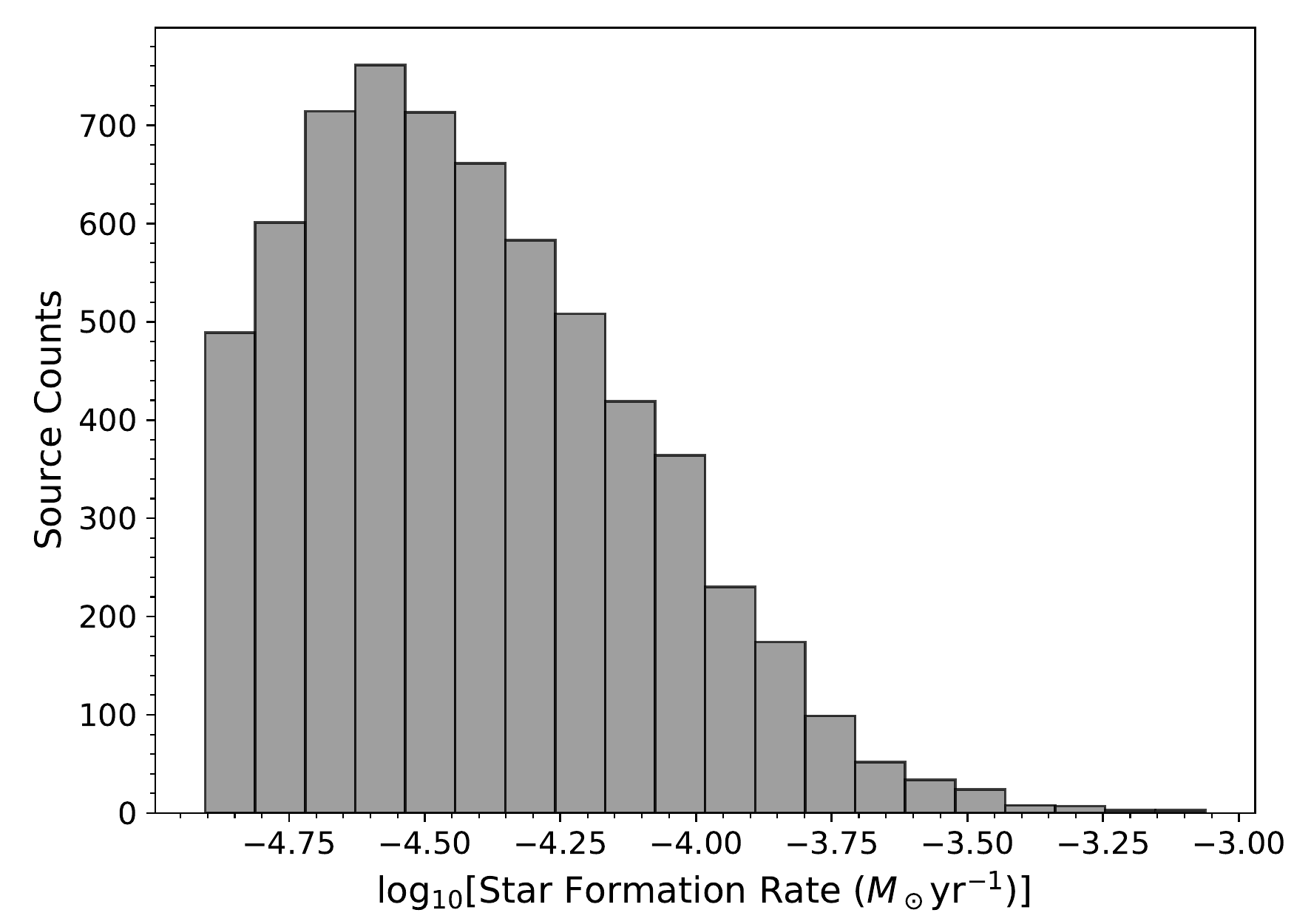}
  \caption{Distribution of star formation rates for ATLASGAL clumps derived using the star formation efficiency determined from the model and assuming an embedded formation time-scale of $2$\,Myr.
  }
  \label{fig:sfr_histo}
\end{figure}

The next question is what fraction of the ATLASGAL clumps are star-forming and the answer to this question is all of them. From a visual inspection of mid- and far-infrared images \citet{urquhart2022} found that only $\sim$10\,per\,cent of the clumps are quiescent and many of these appear to be associated with molecular outflows (\citealt{yang2018, yang2021}), which may be the earliest indication of the formation of a hydrostatic core. The statistical lifetime of massive starless clumps is very short ($< 1-3\times 10^4$\,yr; \citealt{Motte_2018}) and, given the masses and densities of ATLASGAL clumps, it is likely that these will form a cluster in the next 2\,Myr.
Figure\,\ref{fig:sfr_histo} shows the star formation rates for all ATLASGAL clumps with masses over 100\,\msun. Expressing the star formation time-scale in terms of the free-fall timescale at the density of our clumps we can write $\tau_{\rm sf} = f\times \tau_{\rm ff}$, where $f \ge 1.0$. We can now estimate the star formation efficiency per free-fall time using $\varepsilon/f = 1.5\pm 0.76$\,per\,cent, which is slightly above the theoretical value proposed by \citet{krumholz2005}, but in good agreement with previous studies (e.g. 1.8\,per\,cent and 2.6\,per\,cent reported by \citealt{lada2010} and \citealt{pokhrel2021} respectively).

\begin{figure}
  \includegraphics[width=\linewidth]{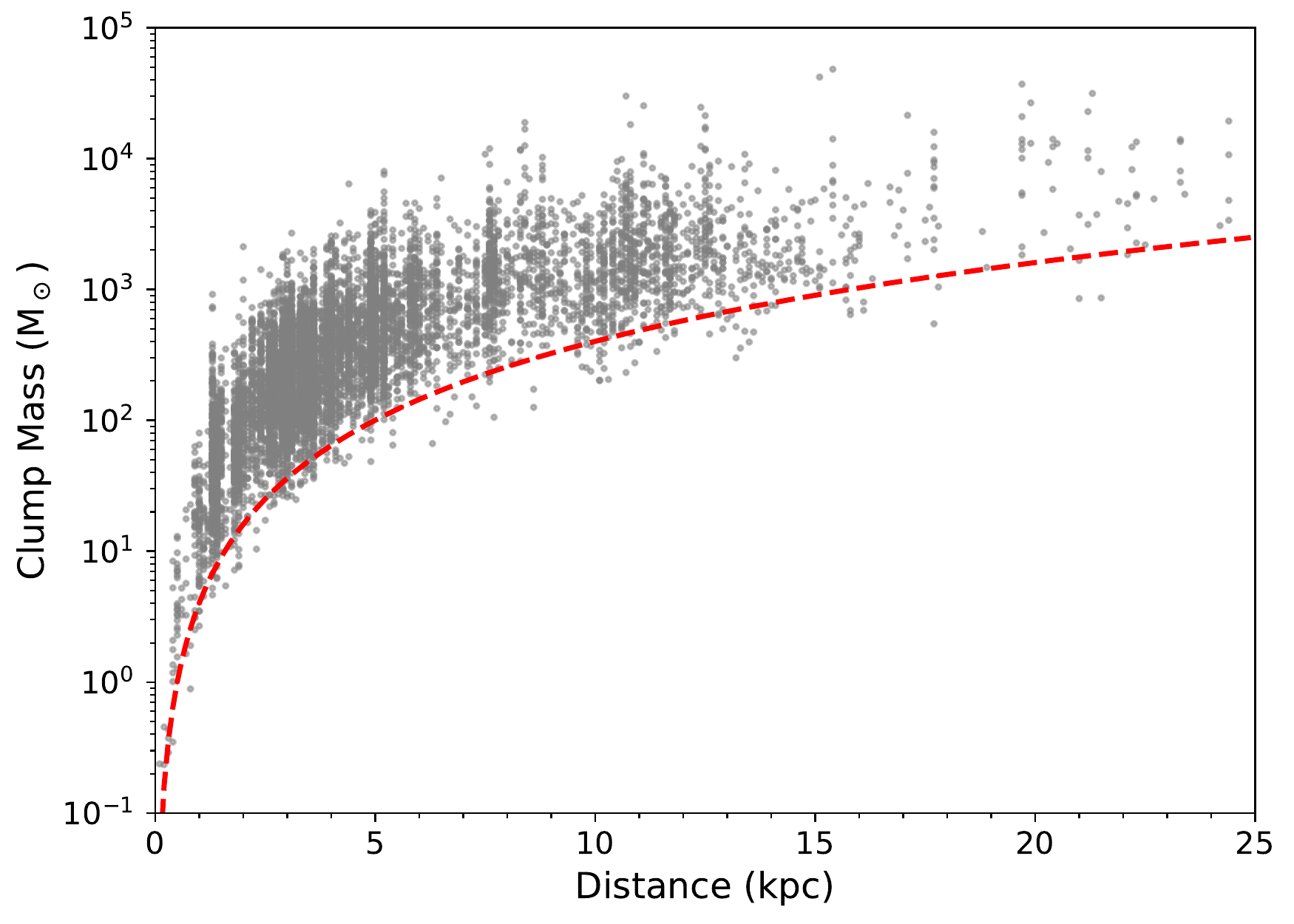}
  \caption{Distribution of clump masses as a function of heliocentric distance. The dashed line shows the mass sensitivity limit for a dust temperature of 25\,K.}
  \label{fig:completeness}
\end{figure}

The next step is to use the SFRs determined for the individual  ATLASGAL clumps to estimate the Galactic star formation rate. However, this is not just a simple process of summing up the contributions from all of the clumps, as we need to take into account the survey's completeness. In Figure\,\ref{fig:completeness} we show the clump mass distribution as a function of distance, which shows that we are complete to all clumps with masses above a thousand solar masses across the disc but are likely to miss a significant number of lower-mass clumps. So to obtain a reliable estimate for the SFR in the inner disc, we need to use a mass threshold and distance limits to select a representative sample that we can be sure is reasonably complete. 

Using Fig.\,\ref{fig:completeness}, we find that we are complete to all clumps with masses above $\sim$100\,\msun\ out to a distance of 6\,kpc. These criteria deliver a large and representative sample of clumps covering a significant fraction of the inner disc (i.e. $2\,{\rm kpc} < R_{\rm gc} < 8.35$\,kpc).  The ATLASGAL catalogue does not provide the physical properties for clumps located towards the Galactic centre ($|\,\ell\,| < 3\degr$) due to difficulties allocating reliable distances to clumps in this region, and this essentially excludes sources within 2\,kpc of the Galactic centre. 

\begin{figure}
  \includegraphics[width=\linewidth]{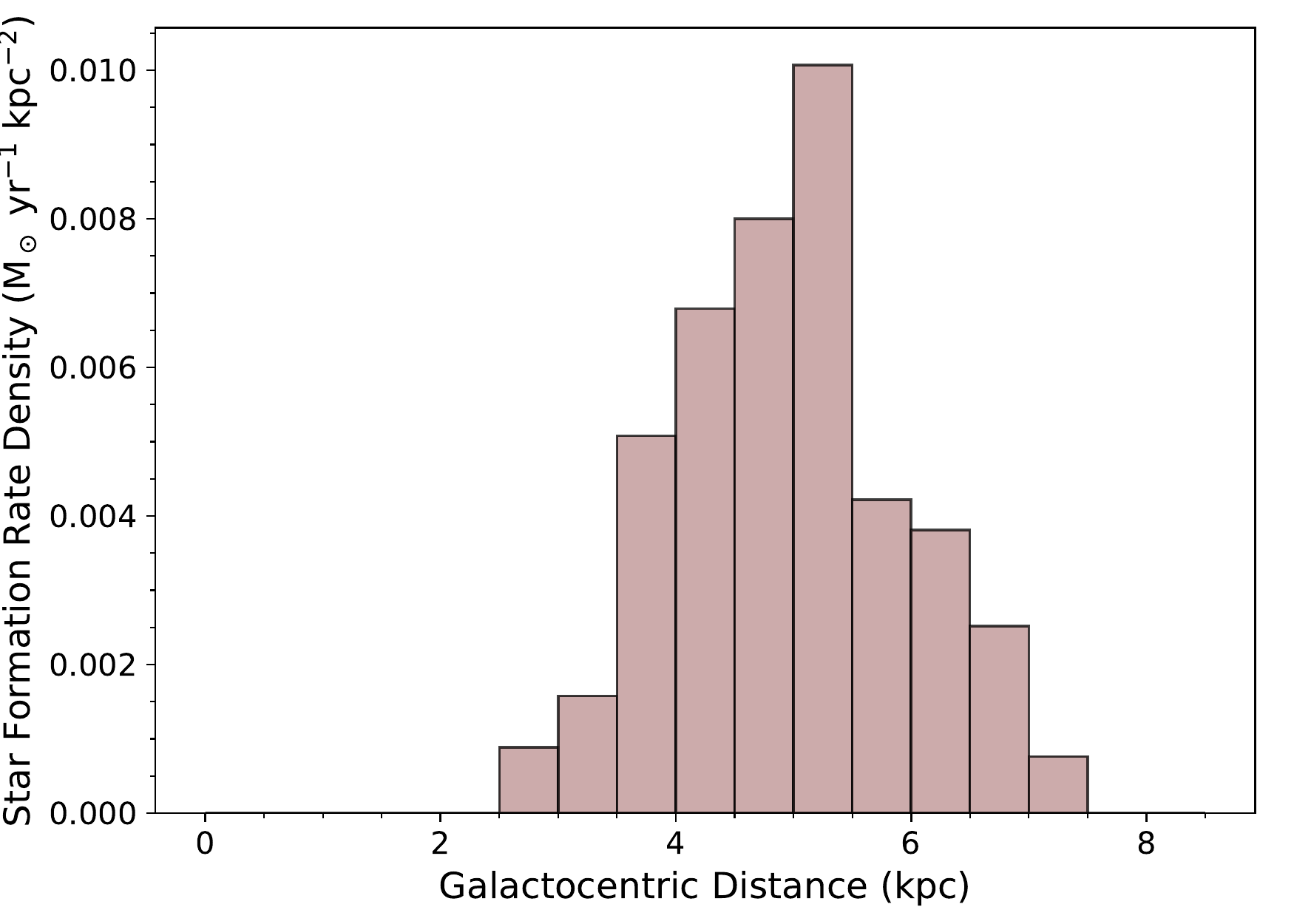}
    \includegraphics[width=\linewidth]{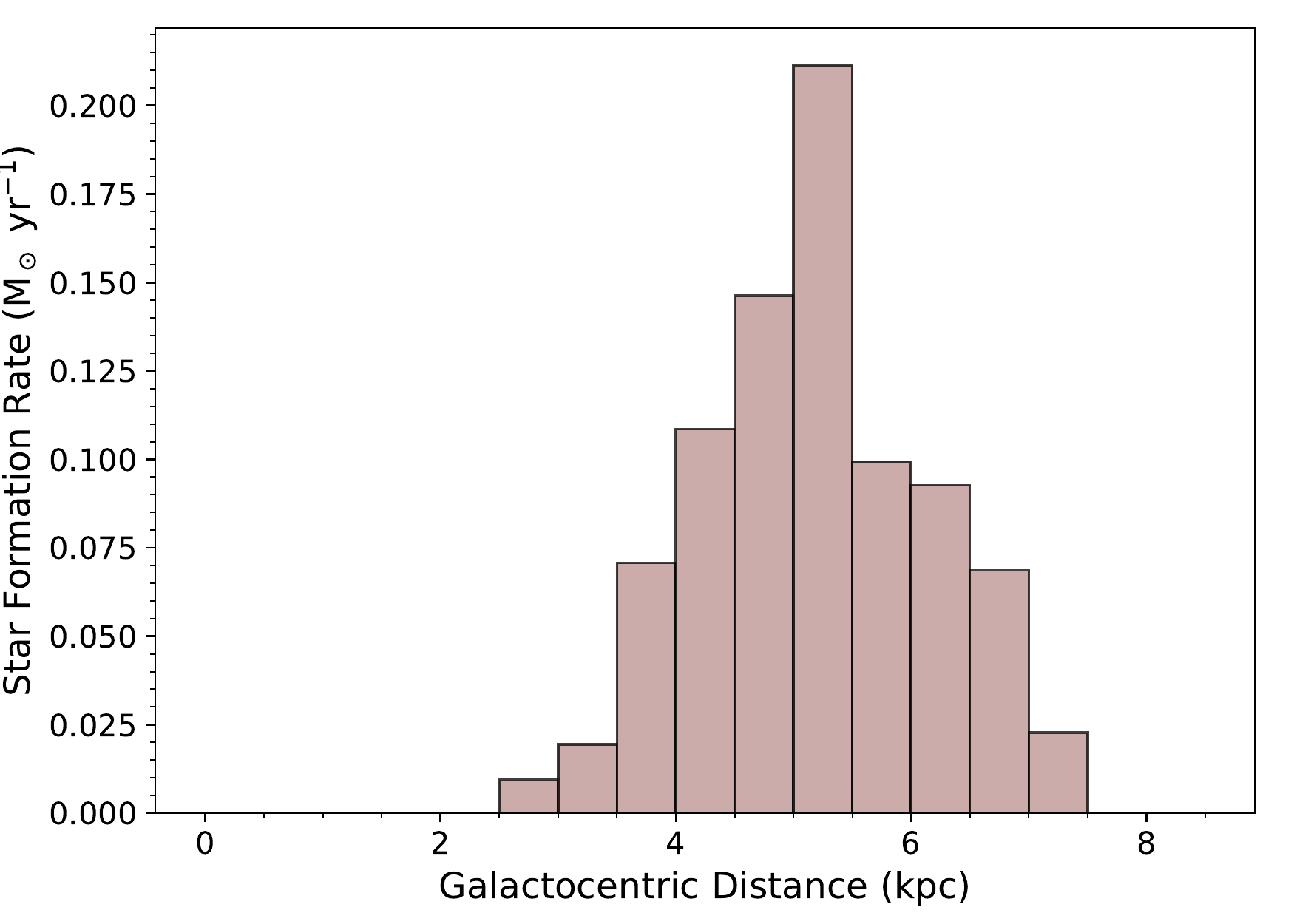}
 \caption{Star formation rate density and star formation rate are shown as a function of Galactocentric radius in the upper and lower panels.  The bin size used is 0.5\,kpc.
 }
  \label{fig:sfr_vs_rgc}
\end{figure}

We separate this sample into Galactocentric bins 0.5\,kpc wide and sum up the SFRs of the clumps in each bin. We then divide this by the area of each of these annuli covered by ATLASGAL at a heliocentric distance of 6\,kpc; this provides a star formation rate density (see upper panel of Fig.\,\ref{fig:sfr_vs_rgc}). The final step is to multiply the SFR density by the area of each annulus to obtain SFR across the inner disc; this is shown in the lower panel of Fig.\,\ref{fig:sfr_vs_rgc}. Summing up the contributions from the individual Galactocentric bins, we obtain a value for the SFR from the inner Galactic disc of 0.68\,\msun\,yr$^{-1}$. This is a lower limit as we have excluded the contribution from clumps below 100\,\msun. 

We can combine this with the star formation taking place in the Galactic centre and the outer Galaxy to obtain an estimate for the total Galactic star formation rate. The SFR in the central molecular zone (CMZ; $|\,\ell\,| < 1\degr$ and $|\,b\,| < 0.5\degr$), where some well known star forming regions are located, such as Sagittarius B1 and B2, G0.6, G0.3 (\citealt{immer2012a, barnes2015})  is $\sim$0.09\,\msun\,yr$^{-1}$ (\citealt{barnes2015}); this value includes some of the most prominent regions in the CMZ but is a lower limit on the star formation taking place in the Galactic centre. Adding this to the value we have for the disc gives a lower limit for the total SFR for the inner Galaxy of 0.77\,\msun\,yr$^{-1}$. According to \citet{miville-deschenes2017}, 85\,per\,cent of the molecular gas is located within the Solar circle (i.e. $R_{\rm gc} < 8.35$\,kpc). If we assume that the distribution of dense gas is similar to that of the molecular gas then it follows that the outer Galaxy will contribute 15\,per\,cent of the total star formation in the Milky Way. Factoring this in, we obtain a lower limit for the Galactic SFR of $0.9\pm 0.45$\,\msun\,yr$^{-1}$. The uncertainty in this measurement is dominated by the uncertainty in the infrared-excess time-scale mentioned earlier.

The Galactic SFR has been investigated by a number of different groups over the years with values reported ranging from 0.35 to 2.6\,\msun\,yr$^{-1}$, with most values lying between 1 and 2\,\msun\,yr$^{-1}$ (see table\,1 of \citealt{2011Chomiuk} for a summary of reported values). \cite{Robitaille_2010} used population synthesis based on data from the {\em Spitzer}/IRAC GLIMPSE survey to determine the GSFR with the result 0.68--1.45\,\msun\,yr$^{-1}$. \cite{davies2011} reported a value of 1.5--2\,\msun\,yr$^{-1}$ from a simulation based on the distribution of Massive YSOs and \hii\ regions characterised by the Red MSX Survey (RMS; \citealt{lumsden2013}). \citet{Licquia_2015} obtained a value of $1.65\pm0.19$\,\msun\,yr$^{-1}$ by combining measurements of properties of the Milky Way and a hierarchical Bayesian statistical method that accounts for the possibility that any value may be incorrect or have underestimated errors. 

Our value of $0.9\pm 0.45$\,\msun\,yr$^{-1}$ compares very well with the range obtained by \citet{Robitaille_2010} but is slightly below those found by the other two studies. 
The average of the three studies is 
$1.48\pm0.28$\,\msun\,yr$^{-1}$, which
is consistent with our result,
given the large uncertainties. 

\section{Discussion}
\label{sect:discussion}

In the previous two sections, we have derived an empirical relationship between clump mass and star formation efficiency and found a trend for decreasing SFE with increasing clump mass. This result is a little unexpected; however, it is consistent with reliable SFE measurements reported in the literature and this provides strong support for our findings. We have used these results to determine a value for the Galactic SFR and find this to be in good agreement with previous studies. 
The reason for the decrease in star formation efficiency for high-mass clumps is unclear from the current data but may be linked to increased feedback from high-mass stars (c.f. \citealt{rugel2019}), which are statistically more likely to be found in the more massive clumps.

In this section we will discuss the assumptions that we have used to determine the form of the star formation efficiency and dense clump mass, which is based on the empirical relation revealed in Fig.\,\ref{fig:lm_diagram_hii}, and how these might impact on our results.

\subsection{Changes in the clump mass}

The masses of clumps may change during the star formation process with material being taken away in the form of molecular outflows and accreted onto stars, while at the same time new material is infalling onto the clump from the molecular cloud in which the clumps reside. The fraction of the initial mass that is converted into stars has been taken into account when calculating the star formation efficiency (Eqn.\,\ref{eq:sfe1}), however, changes due to infall and outflows have not yet been considered.

Modelling by \citet{Machida_2012} found that 8-49\,per\,cent of the initial core mass can be ejected back into the interstellar medium through outflows, with 26--54\,per\,cent going into stars. If we assume that a similar amount of material goes into both stars and outflows in cores, we can put an upper limit on the mass lost through outflows. Only a relatively small fraction of the clump mass is converted into stars ($\varepsilon \approx 0.2$) and so we might expect the clump to lose a further $\sim20$\,per\,cent of its mass during the star formation process. 

Infall rates for high-mass star-forming clumps are between 0.3 to $16 \times 10^{-3}$\,\msun\,yr$^{-1}$ (e.g., \citealt{wyrowski2016}). If we take the time-scale for high-mass star formation ($\sim5\times 10^5$\,yr; \citealt{davies2011, mottram2011b}) and assume that the infall continues at the same rate over this time-scale, these rates correspond to total accumulation of 150-8000\,\msun. Infalling material therefore has the potential to have a larger impact on the clump mass over time than the mass loss due to outflows. However, comparisons of clump masses as a function of the level of star formation in the ATLASGAL sample revealed no significant changes as a function of evolution (\citealt{urquhart2022}), which would suggest that the material lost via outflows and accretion onto stars is roughly balanced by new material falling onto clumps during the star formation process or that the rates are at the low end of the ranges.

\subsection{Luminosities of embedded clusters}

We have used the bolometric luminosity of the \hii\ regions to derive our empirical relationships; however, is this a reliable measure of a cluster's luminosity, given that many of the lower-mass stars have yet to contract down onto the main sequence?  Low-mass stars are actually more luminous in their pre-main-sequence stages (a factor of $\sim$2 higher) and become less luminous as they contract and move towards the main sequence. This will lead to the masses of lower-mass stars being overestimated and resulting in a slight overestimation of the SFE. However, the average mass of a star in a cluster is 0.5\,\msun\ and so, given the relationship between mass and luminosity is $L \propto M^{4}$ for main-sequence stars below 1\,\msun, the contribution to the luminosity for these lower-mass cluster members is very small and can be safely ignored.

Another consideration is the contribution to the observed luminosity from accretion. The total luminosity is given by:

\begin{equation}
L_{\rm bol} = L_\star + L_{\rm acc}
\end{equation}

\noindent where $L_\star$ and  $L_{\rm acc}$  are the stellar luminosity and the accretion luminosity, respectively. We estimate the latter using the simplified relation between  accretion luminosity and accretion rate \citep{wolfire1987}:

\begin{equation}
L_{\rm acc} = \frac{G \dot M M_\star}{R_\star},
\end{equation}

\noindent where $M_\star$ and $R_\star$ are the mass and radius of the star and $\dot M$ is the accretion rate. Mass accretion rates in massive dense cores range between 10$^{-4}$ and 10$^{-2}$\,\msun\,yr$^{-1}$ (\citealt{Motte_2018} and references therein). 
The size of the embedded protostar  can vary between 10 and 100\,R$_\odot$ (\citealt{hosokawa2009}) and is therefore poorly constrained. Assuming an accretion rate of $\dot M = 10^{-3}$\,\msun\,yr$^{-1}$ and a protostellar size of 30\,R$_\odot$ we estimate the accretion luminosity to be approximately $3\times 10^4$\,\lsun\ for a 20\,\msun\ protostar; this is comparable to the stellar luminosity of a main-sequence star of the same mass. However, the outflow studies appear to show that the outflow strength decreases with time and, given that these are linked to accretion, this suggests that the accretion luminosity will also decrease over time (\citealt{de_villiers2015,lopez2011, motte2007}) and so this can be considered to be an upper limit. Indeed the models by \citet{hosokawa2009} would suggest that once a star has reached a mass of $\sim$10\,\msun\ the stellar luminosity is approximately 5 times larger than the accretion luminosity.

\begin{figure*}
    \centering
    \includegraphics[width = 0.49\textwidth]{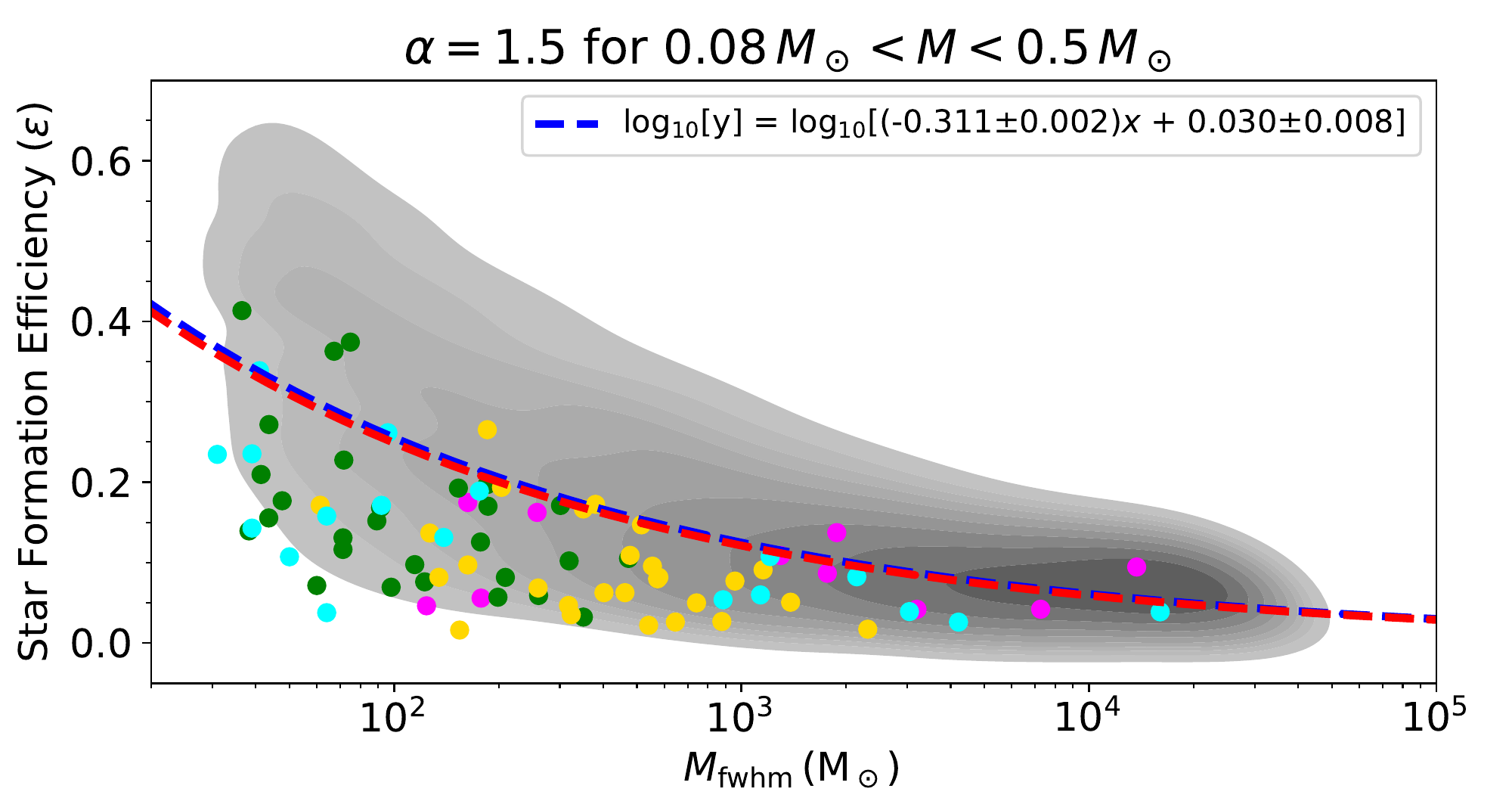}
     \includegraphics[width = 0.49\textwidth]{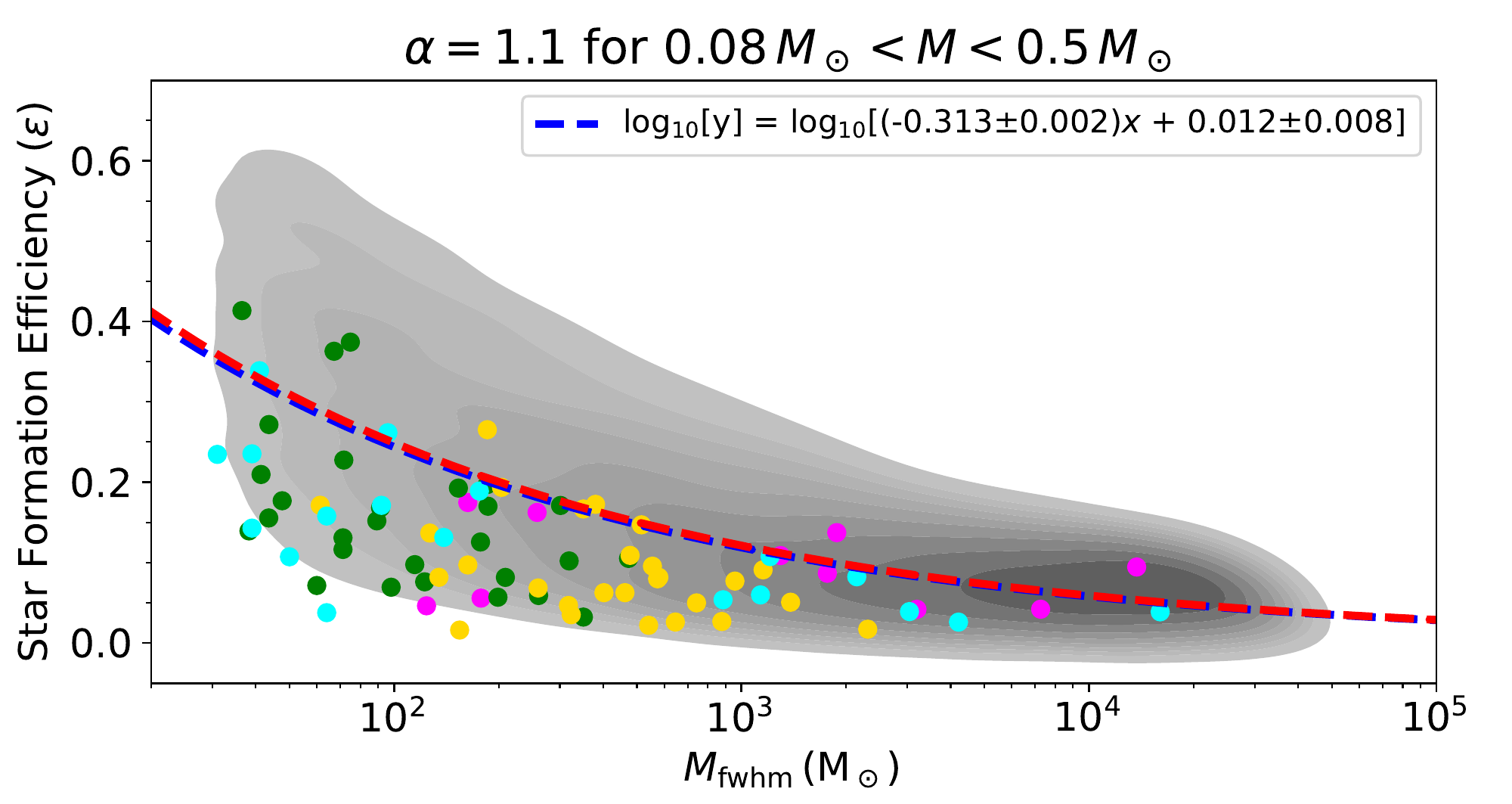}
       \includegraphics[width = 0.49\textwidth]{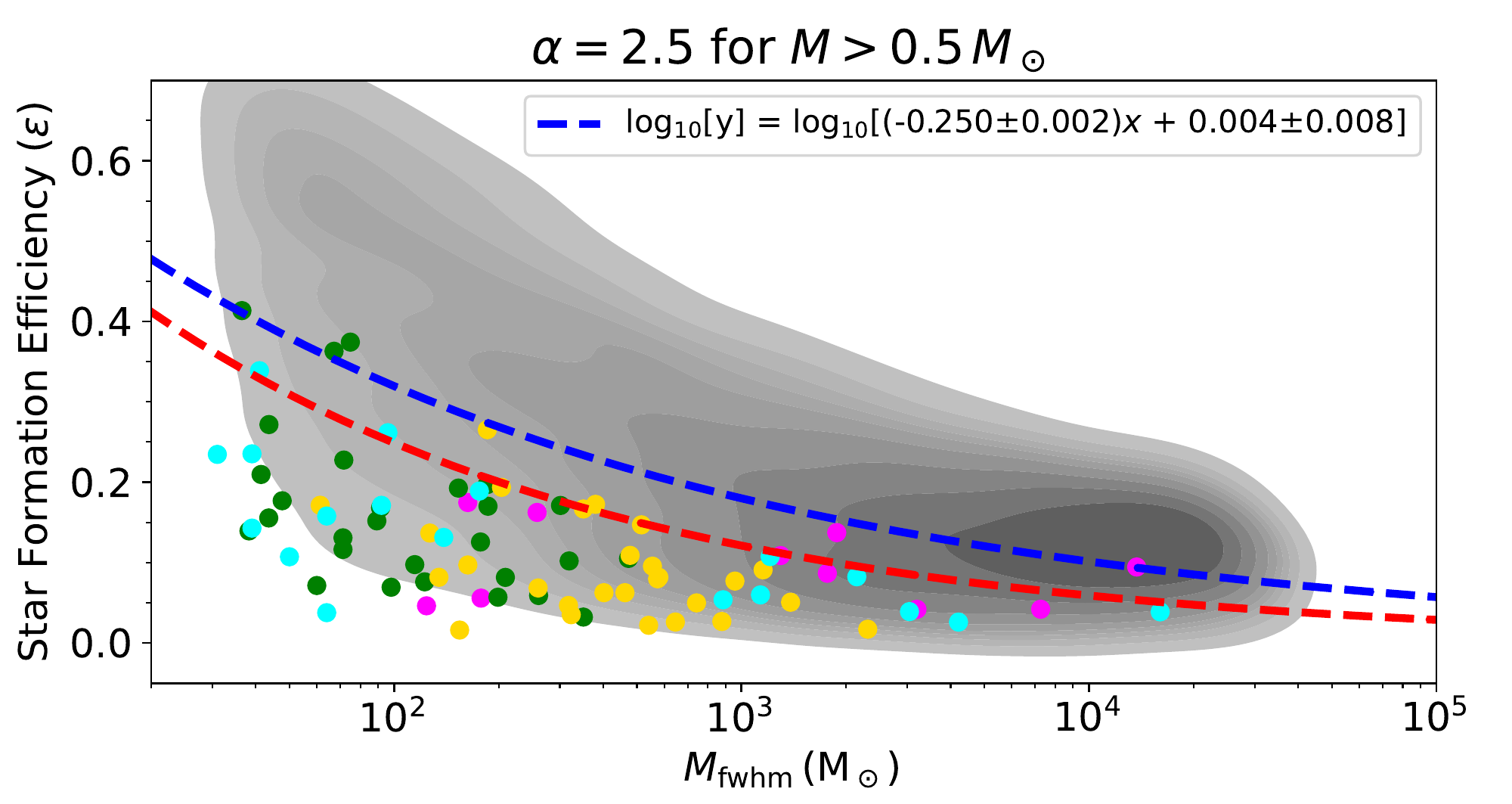}
     \includegraphics[width = 0.49\textwidth]{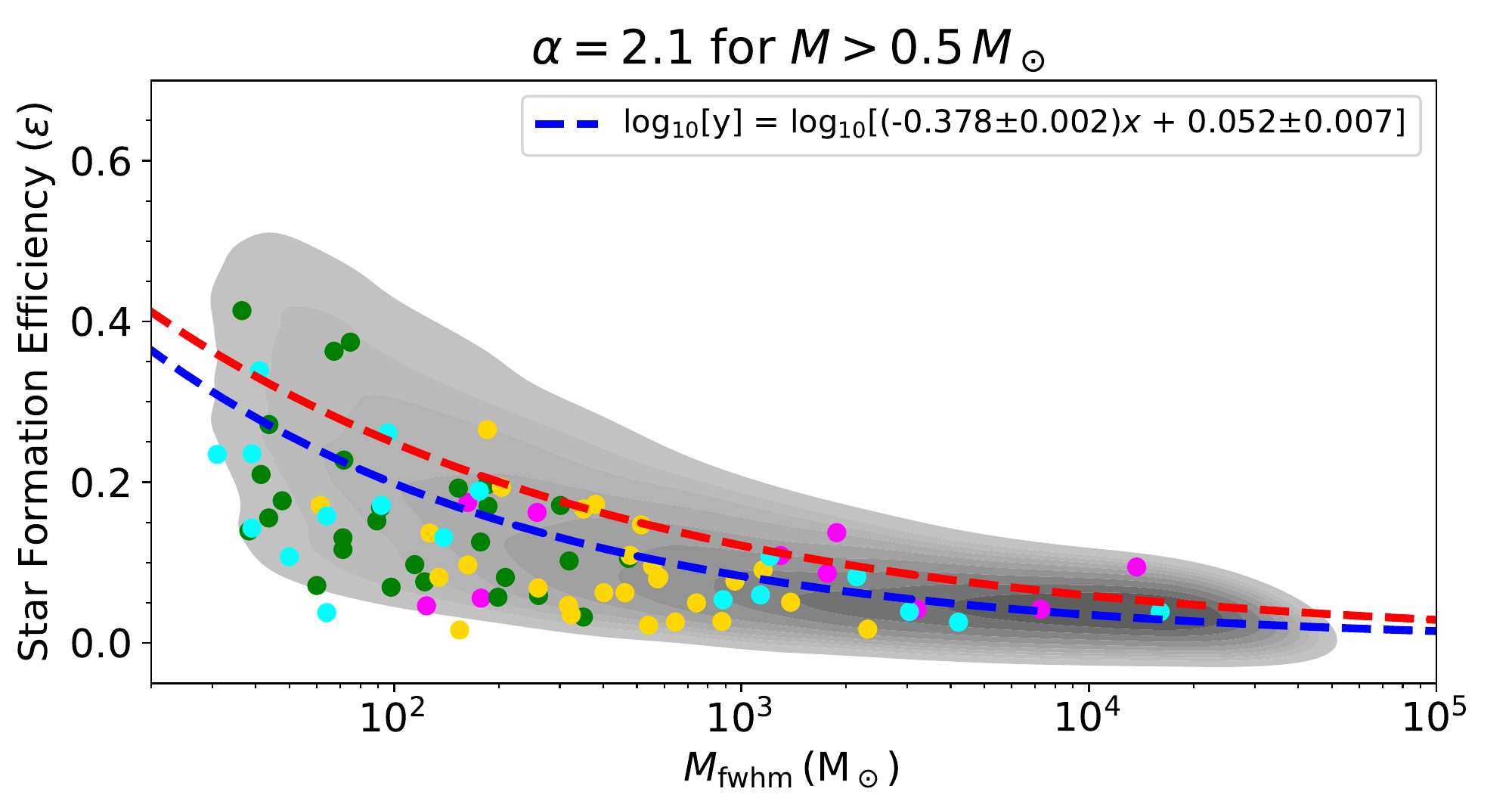}

    \caption{Distribution of star formation efficiencies obtained from the Monte Carlo code as a function of clump mass for various IMFs. The contours and filled circles are as defined in Fig.\,\ref{fig:sfe_dist_map}. The red curve is the fit to the relationship obtained from the standard IMF and given in Eqn.\,\ref{eq:sfe} while the green curve is the fit to the results shown. The title of each plot indicates the parameter that has been changed. } 
    \label{fig:sfe_varying_imf}
    
\end{figure*} 

The accretion rates for lower-mass protostars forming in the cluster are 2--3 orders of magnitude lower than for the high-mass protostars ($\sim10^{-6}$-10$^{-5}$\,\msun\,yr$^{-1}$; \citealt{Rygl2013}) and, even though they are much more numerous than high mass protostars, their contribution to the overall accretion luminosity is relatively modest ($\sim$10\,per\,cent). 

In the worst case scenario the stellar luminosity of the clusters will have been overestimated by up to a factor of two, leading to an overestimate of the star formation efficiency by a similar amount; however, in most cases it is likely to be much smaller and so this is unlikely to affect our results.

\subsection{Variations in the IMF}
\label{sect:IMF_variations}

We have used the standard \citet{kroupa2001} IMF to determine the star formation efficiencies in this study; however, there is increasing evidence that the IMF is not as universal as first thought (e.g., \citealt{dib2017}) and so we have investigated how modest changes to the shape of the adopted IMF affects the derived SFEs. In Figure\,\ref{fig:sfe_varying_imf} we show the result obtained from changing the $\alpha$ by $\pm$0.2 for 0.08\,\msun\, $\le M \le $ 0.5\,\msun\ (upper panels) and $\alpha$ by $\pm$0.2 for $ M > $ 0.5\,\msun\ (lower panels). 

Changing the slope of the lower-mass stars has no significant impact on the results. Changing the slope of the higher mass stars has a significant impact on the star formation efficiency. A steeper slope (lower left panel of Fig.\,\ref{fig:sfe_varying_imf}) will result in fewer higher mass stars (bottom-heavy IMF) and allow more lower-mass stars to be added to the cluster before the $L/M$ threshold is breached, resulting in a high star formation efficiency (an increase of $\sim$0.05--0.1). However, this results in a significantly poorer fit to the observational SFE measurements and is therefore be considered less likely. 

The results for the shallower slope for the higher mass stars (top heavy) are shown in the lower right panel of Fig.\,\ref{fig:sfe_varying_imf}. This plot shows overall lower values for the SFE than our previous results by 0.05 and arguably a better fit to the observational values. This is due to the presence of a higher number of more massive stars in the clusters, leading to the $L/M$ threshold being breached earlier than otherwise expected and resulting in the formation of lower-mass clusters.  Although the decrease in the SFE appears relatively modest, it corresponds to an overall decrease of $\sim$25\,per\,cent and as such will have a significant impact of the estimate of the Galactic star formation rate, reducing it to $\sim0.68\pm 0.34$\,\msun\,yr$^{-1}$, which is significantly lower than the mean value of $\sim1.48$\,\msun\,yr$^{-1}$ determined by the three studies discussed earlier. 

These tests show that the shape of the IMF for lower-mass stars have little influence on the resulting star formation efficiency of a cluster. However, increasing and decreasing the slope of the higher-mass stars ($>0.5$\,\msun) does have a significant impact of the SFE, with a steeper slope increasing the SFE by 0.05 and a shallower slope decreasing it by a similar amount. However, the fit to the observational data is poor for the steeper slope and although the shallower slope results a better fit with the observed data the corresponding Galactic star formation rate that is significantly below previously determined values. From this simple analysis we conclude that large variations in the IMF, although possible, are not a good fit to the available data.


\section{Conclusions}
\label{sect:conclusions}

We have used an empirical $L/M$-ratio derived from a fit to \HII\ region properties and a Monte Carlo simulation to determine a relationship between a clump's star formation efficiency and its dense gas mass. We have used these values to estimate the Galactic star formation rate. Our main findings are as follows:

\begin{itemize}
    \item The average star formation efficiency of $0.17$ with a standard deviation of 0.04. We find the star formation efficiency decreases as a function of clump mass with values of $\sim$0.25 for mass of a few hundred solar masses dropping to $\sim$0.08 for masses of a few thousand solar masses. The lower star formation efficiency for high mass clumps may be due to the increased feedback from high-mass stars that are significantly more likely to form in them.\\
    
    \item We find good agreement between the SFE obtained from our model and observational measurements based on YSO source counts in nearby molecular clouds, with 90\,per\,cent of the measurements agreeing with the model predictions. The observational measurements are considered to be the most reliable available and so the correlation between the predicted values and observations provides strong support for the model. \\
    
    \item We use the star formation efficiencies to predict the total mass of dense gas in the Galactic disc that will be turned into stars and, using the infrared excess time of $2\pm1$\,Myr as the embedded YSO time-scale, we estimate the Galactic star formation rate to be approximately $0.9\pm0.45$\,\msun\,yr$^{-1}$.\\
    
    \item Although our calculations are derived from a few simple assumptions, the star formation efficiencies and Galactic star formation rate are in good agreement with previously published values, which provides strong support for the SFE obtained from our model.

\end{itemize}

This model has proven to be reliable and has an advantage over previous studies by avoids the uncertainty associated with converting the infrared luminosity to stellar mass that affects many of the published extragalactic and Galactic studies.

\section*{Acknowledgements}

We would like to thank the referee for their constructive comments and suggestions that have helped put this work in a broader context. We thank Alex Pettitt for providing his help preparing Fig.\,\ref{fig:atlasgal} and Dirk Froebrich, Timothy Kinnear  and Robert Gutermuth for insightful discussions that have helped improve the clarity of this work and to Riwaj Pokhrel for providing an electronic versions of the tables from his 2020 paper. This research made use of Astropy,\footnote{http://www.astropy.org} a community-developed core Python package for Astronomy \citep{astropy2013, astropy2018}; matplotlib \citep{matplotlib2007}; numpy and scipy \citep{scipy2020}. This  research  has  made  use of  the  SIMBAD  database  and  the  VizieR  catalogue,  operated  at  CDS,  Strasbourg, France. This document was prepared using the Overleaf web application, which can be found at www.overleaf.com.

\section*{Data Availability}

The data presented in this article is available from a dedicated website:
\url{https://atlasgal.mpifr-bonn.mpg.de/}




\input{SFE_of_Clusters_2022.bbl}

\bsp	
\label{lastpage}
\end{document}